\documentclass[sigconf]{acmart}
\usepackage{subfigure}
\usepackage{graphicx}
\usepackage{amsmath}
\usepackage{bm}
\usepackage{url}
\usepackage{stmaryrd}
\usepackage{balance}
\usepackage{amssymb}
\usepackage{pgfplots}
\usepackage{pifont}
\usepackage{epsfig}
\usepackage{booktabs}
\usepackage{pgffor}
\usepackage{tikz}
\usepackage{multirow}
\usepackage{algorithm}
\usepackage{algorithmic}
\definecolor{purple}{rgb}{0.6,0.2,0.8}

\usepackage[para]{footmisc}

\makeatletter
\newcommand\xleftrightarrow[2][]{%
  \ext@arrow 9999{\longleftrightarrowfill@}{#1}{#2}}
\newcommand\longleftrightarrowfill@{%
  \arrowfill@\leftarrow\relbar\rightarrow}
\makeatother

\newcommand{\mb}{\mathbf}

\newcommand{\problem}[0]{\textsc{Miso-Klic}}

\newcommand{\our}[0]{\textsc{IDEA}}

\newcommand{\chinesematch}[0]{\textsc{CT-EM}}
\newcommand{\englishmatch}[0]{\textsc{ET-EM}}
\newcommand{\chinesescore}[0]{\textsc{CT-S}}
\newcommand{\englishscore}[0]{\textsc{ET-S}}

\setcopyright{rightsretained}

\begin{document}

\fancyhead{}

\title{Missing Movie Synergistic Completion across Multiple Isomeric Online Movie Knowledge Libraries}


\author{
Bowen Dong$^\star$, Jiawei Zhang$^\P$, Chenwei Zhang$^\star$, Yang Yang$^\dagger$ and Philip S. Yu$^{\star \ddagger}$}

\affiliation{{$^\P$IFM Lab, Florida State University, Tallahassee, FL, USA}\\ 
		{$^\star$University of Illinois at Chicago, Chicago, IL, USA}\\ 
		{$^\dagger$Beihang University, Beijing, China}
      }
\email{jiawei@ifmlab.org, [czhang99, bdong5, psyu]@uic.edu, yangyangfuture@hotmail.com}

\begin{abstract}

Online knowledge libraries refer to the online data warehouses that systematically organize and categorize the knowledge-based information about different kinds of concepts and entities. In the era of big data, the setup of online knowledge libraries is an extremely challenging and laborious task, in terms of efforts, time and expense required in the completion of knowledge entities. Especially nowadays, a large number of new knowledge entities, like movies, are keeping on being produced and coming out at a continuously accelerating speed, which renders the knowledge library setup and completion problem more difficult to resolve manually. In this paper, we will take the online movie knowledge libraries as an example, and study the ``\textit{\underline{M}ultiple aligned \underline{IS}omeric \underline{O}nline \underline{K}nowledge \underline{LI}braries \underline{C}ompletion problem}'' ({\problem}) problem across multiple online knowledge libraries. {\problem} aims at identifying the missing entities for multiple knowledge libraries synergistically and ranking them for editing based on certain ranking criteria. To solve the problem, a thorough investigation of two isomeric online knowledge libraries, Douban and IMDB, have been carried out in this paper. Based on analyses results, a novel deep online knowledge library completion framework ``\textit{\underline{I}ntegrated \underline{D}eep align\underline{E}d \underline{A}uto-encoder}'' ({\our}) is introduced to solve the problem. By projecting the entities from multiple isomeric knowledge libraries to a shared feature space, IDEA solves the {\problem} problem via three steps: (1) entity feature space unification via embedding, (2) knowledge library fusion based missing entity identification, and (3) missing entity ranking. Extensive experiments done on the real-world online knowledge library dataset have demonstrated the effectiveness of {\our} in addressing the problem.

\end{abstract}



\maketitle

\section{Introduction}\label{sec:intro}

Online knowledge libraries refer to the online data warehouses that systematically capture, organize, and categorize the knowledge-based information about different kinds of information concepts and entities. Depending on the categories of the concepts covered, the online knowledge libraries come in two forms:

\begin{itemize}
\item \textit{General Online Knowledge Library}: An online knowledge library involving diverse information about various information topics and concepts. A representative example of \textit{general online knowledge library} is Wikipedia. The knowledge concepts covered in Wikipedia come from very diverse areas, like Culture and Arts, Geography and Places, Society and Social Sciences, History and Events, Natural and Physical Sciences, etc.

\item \textit{Specific Online Knowledge Library}: An online knowledge library containing information about one certain topic or concept. The representative examples of \textit{special online knowledge library} include IMDB (an online knowledge library about movies, TV shows and other videos), Yelp (an online knowledge library about offline restaurants, shops and other POIs (points of interest)), and Stack Overflow (an online knowledge library about programming problems and solutions), etc.
\end{itemize}

In this paper, we will use two \textit{specific online knowledge libraries} both about movies/TV shows as an example to illustrate the problem and solutions \cite{DBLP:journals/corr/abs-1810-10175}. These two \textit{specific online knowledge libraries} studied in this paper include Douban Movie\footnote{https://movie.douban.com} and IMDB Movie\footnote{http://www.imdb.com}. Douban is one of the most influential web 2.0 websites in China providing services about movies, books, music, and recent offline events. As one of the most successful service branch of Douban, Douban Movie provides comprehensive knowledge about the latest and past movies across the world together with the user reviews. IMDB (short for Internet Movie DataBase) is an online database of information related to movies, television programs and video games. The information about the movies/videos/shows covered in IMDB is very diverse, which include cast members, production crew, fictional characters, biographies, plot summaries, trailers and review ratings. Specifically, in this paper, we will mainly focus on the \textit{movies} covered in Douban and IMDB only. For simplicity reasons, we will misuse ``Douban'' and ``IMDB'' in referring to ``Douban Movie'' and ``IMDB Movie'' in the following parts of this paper.

For most of the movies, both Douban and IMDB will create a specific homepage introducing their basic information, like the storyline, genres, on-show date, production country, and casts involved in the movies. What's more, Douban and IMDB will also provide the pages for the audiences and critics to post reviews and ratings about the movies. Based on the information provided in Douban and IMDB, people can generally get both the official information about the movies released by the movie producers and the audiences' personal preferences and watching experiences regarding the movies. 

Even though Douban and IMDB are both movie-related \textit{online knowledge libraries}, they have very significant differences in both representation language and content due to their different oriented users and market.

\begin{itemize}
\item \textit{Representation Language}: Douban is created to provide movie related services for the Chinese users specifically, and the language used in Douban is Chinese. Meanwhile, IMDB is a US oriented and worldwide accessible movie site, and the language used in IMDB is English.

\item \textit{Individual Generated Content}: Due to the differences in the market orientation and users, the content generated in Douban, like reviews, mainly indicates the Chinese users' opinions regarding the movies, while the content generated in IMDB denotes the US/worldwide users' attitudes towards the movies.
\end{itemize}
Formally, we name the knowledge libraries containing common information entities but with totally different representation languages and contents as the ``\textit{aligned isomeric online knowledge libraries}'', whose formal definition will be provided in Section~\ref{sec:formulation} in detail.

Generally, the setup of \textit{online knowledge libraries} is an extremely challenging and laborious task, in terms of both labor, time and expense required in the completion of knowledge entities. What's more, new knowledge entities, like movies, are keeping on being produced and coming out at a continuously accelerating speed, which renders the knowledge library completion problem more difficult to resolve manually. For most online knowledge libraries, automatically identifying the missing entities (e.g., movies in our cases) which already exist in other knowledge libraries is a fundamental problem for the service providers \cite{WWZL16}. Meanwhile, among the identified missing entities, not all of them should be created immediately due to the limited work forces and other issues (e.g., some entities may be forbidden or not welcome in certain countries). Providing a reasonable ranking of the identified missing entities, where those can make up the greatest knowledge gap in the knowledge libraries should be ranked at the top, will be a great way to solve the problem.

\noindent \textbf{Problem Studied}: In this paper, we will study the \textit{\underline{M}ultiple aligned \underline{IS}omeric \underline{O}nline \underline{K}nowledge \underline{LI}braries \underline{C}ompletion} ({\problem}) problem, which aims at both identifying and ranking the missing information entities for each of the knowledge library synergistically. To the best of our knowledge, we are the first to propose the \textit{multiple aligned isomeric online knowledge libraries} concept and we are also the first to study the knowledge library completion problem across \textit{multiple aligned isomeric online knowledge libraries} simultaneously.

Novel research problems always come along with great challenges, since few prior research works are available for reference and majority of the problem remain to be open. The challenges covered in the {\problem} problem come in several different aspects:
\begin{itemize}

\item \textit{Concept Definition and Problem Formulation}: Both the \textit{multiple isomeric online knowledge libraries} concept and the {\problem} problem studied in this paper are novel, which are proposed in this paper for the first time. Formal definitions of both the concepts and problem are the prerequisites for solving the problem.

\item \textit{Online Knowledge Library Statistical Analyses}: Few works have been done on analyzing \textit{multiple isomeric online knowledge libraries} before. Besides the obvious differences in presentation languages and contents generated by individual users, more thorough and comprehensive statistical analyses need to be performed to reveal other hidden differences.

\item \textit{Isomerism of Online Knowledge Libraries}: The isomeric differences of the online knowledge libraries, e.g., the presentation language and content differences, will introduce a number of obstacles in the synergistic completion of multiple online knowledge libraries. A powerful model which can handle such isomeric data in totally different languages will be desired.

\end{itemize}

To resolve these challenges aforementioned, in this paper, we will first introduce the formal definitions of the \textit{multiple isomeric online knowledge libraries} concept and the {\problem} problem. Based on the concept and problem definitions, we will analyze the online knowledge library datasets thoroughly to study the concrete differences, which will provide us with the insights to design the model to solve the {\problem} problem. According to the analyses results, a novel deep online knowledge library completion framework, namely ``\underline{I}ntegrated \underline{D}eep align\underline{E}d \underline{A}uto-encoder'' ({\our}), is introduced in this paper, which is based on the deep auto-decoder model. Framework {\our} involves three steps: (1) entity feature space unification via emebdding, (2) entity fusion based on alignment, and (3) missing entity ranking. In the first step, model {\our} effectively extracts the hidden feature representation for the entities from different knowledge libraries in a shared feature space. In the second step, information entities are fused across different libraries with a supervised learning setting, where the isolated entities failing to be aligned will be identified as the missing entities. In the third step of {\our}, these identified missing entities will be ranked synergistically in terms of their importance in bridging the knowledge gaps in different libraries.

The remaining parts of this paper are organized as follows. In Section~\ref{sec:formulation}, we will provide the formal definition of several important concepts used in this paper as well as the formulation of the {\problem} problem. Detailed analyses about the information distributions of Douban and IMDB will be provided in Section~\ref{sec:general_analysis}. The introduction to the deep framework {\our} is available in Section~\ref{sec:method}, whose effectiveness will be evaluated in Section~\ref{sec:experiment} with experiments on real-world datasets. Finally, in Section~\ref{sec:related_works}, we will introduce the existing related works and conclude the paper in Section~\ref{sec:conclusion}.

\section{Terminology Definition and Problem Formulation} \label{sec:formulation}

This paper is based on two real-world online knowledge libraries about movies, Douban and IMDB. In this section, we will first introduce the formal definitions of several important concepts used in this paper and provide the formulation of the {\problem} problem.

\subsection{Concept Definition}

In this paper, our research and analyses works are mainly about the \textit{online movie knowledge libraries} (OMKLs), whose formal definition is provided as follows.

\noindent \textbf{Definition 1} (Online Movie Knowledge Library): An \textit{online movie knowledge library} can be represented as a graph $G = (\mathcal{M}, \mathcal{U}, \mathcal{E}, \mathcal{A})$, where node set $\mathcal{M} = \{m_1, m_2, \cdots, m_n\}$ denotes the set of $n$ movies in the library and $\mathcal{U} = \{u_1, u_2, \cdots, u_l\}$ is the set of $l$ users. Link set $\mathcal{E}$ represents the review/rating records posted by users for certain movies in the library. For instance, link $(u_i, m_j) \in \mathcal{E}$ denotes the review/rating record posted by user $u_i$ on movie $m_j$. Both the nodes and links in $G$ are attached with a set of attributes, which are formally represented as set $\mathcal{A} = \mathcal{A}_n \cup \mathcal{A}_l$. Sets $\mathcal{A}_n$ and $\mathcal{A}_l$ denote the attributes about nodes (both movies and users) and links respectively.

For the datasets to be introduced in this paper, the attributes associated with the movies are very diverse, including \textit{movie name}, \textit{movie textual storeline}, \textit{movie cast} (including \textit{directors}, \textit{writers} and \textit{cators} and \textit{actress}),  \textit{movie genre}, \textit{production country}, \textit{language}, \textit{on-show date}, \textit{length}, etc. Meanwhile, the attribute about the users is relatively simple, which merely contains the \textit{user names} and some basic profile information. For the review/rating links in the OMKLs, the associated attributes involve the \textit{numerical rating}, \textit{textual content} and \textit{timestamp} of the review/rating records. More detailed information about the datasets will be introduced in Section~\ref{sec:general_analysis}.

On the web, multiple OMKLs can exist at the same time, like Douban and IMDB, which can share lots of common movies but with contents in totally different languages. Formally, we name the OMKLs from totally different domains but sharing common information entities as \textit{multiple aligned isomeric online movie knowledge libraries} (or \textit{multiple aligned isomeric OMKLs}).

\noindent \textbf{Definitinon 2} (Multiple Aligned Isomeric OMKLs): Formally, the $k$ \textit{aligned isomeric OMKLs} sharing information about common movie entities can be formally represented as $\mathcal{G} = ((G^{(1)}, G^{(2)}, \cdots, G^{(k)}), \\ (\mathcal{A}^{(1,2)}, \mathcal{A}^{(1,3)}, \cdots, \mathcal{A}^{(k-1,k)}))$, where $G^{(i)}$ represents the $i_{th}$ ($i \in \{1, 2, \cdots, k\}$) OMKL and $\mathcal{A}^{(i,j)}$ ($i,j \in \{1, 2, \cdots, k\}$) denotes the set of undirected \textit{movie anchor links} between OMKLs $G^{(i)}$ and $G^{(j)}$.

In the above definition, the concept \textit{movie anchor link} indicates the correspondence relationships of the same movies between different OMKLs.

\noindent \textbf{Definition 3} (Movie Anchor Link): Given two OMKLs $G^{(i)}$ and $G^{(j)}$, the set of \textit{movie anchor links} between them can be represented as $\mathcal{A}^{(i,j)} \subset \mathcal{M}^{(i)} \times \mathcal{M}^{(j)}$. Pair $(m^{(i)}_p, m^{(j)}_q) \in \mathcal{A}^{(i,j)}$ iff $m^{(i)}_p$ and $m^{(j)}_q$ are actually the same movie in $G^{(i)}$ and $G^{(j)}$ respectively. To represent the corresponding movies of $m_p^{(i)}$ in OMKL $G^{(j)}$ (i.e., ), we can use notation $\mathcal{A}^{(i,j)}(m^{(i)}_p)$ to denote $m^{(j)}_q$ equivalently.

Movie anchor links between pairwise OMKLs, e.g., $G^{(i)}$ and $G^{(j)}$, are undirected. For instance, in referring to the movie anchor links between OMKLs $G^{(i)}$ and $G^{(j)}$, we will use the notations $\mathcal{A}^{(i,j)}$ and $\mathcal{A}^{(j,i)}$ interchangeably in this paper. What's more, for most OMKLs, each movie will have one exact homepage created, and for the movie series, each season/episode is treated as a separate movie in our datasets. Therefore, the movie anchor links between pairwise OMKLs will have an inherent \textit{one-to-one} cardinality constraint.

\subsection{Problem Formulation}

Based on the concept definitions introduced above, we will provide the formulation of the {\problem} problem in this part. Formally, the {\problem} problem involves two sub-problems simultaneously: (1) missing movie identification problem, and (2) missing movie ranking problem, where the missing movies to be ranked in the second sub-problem is based on the result of the first sub-problem. To simplify the statement about the problem setting, in this work, we will take a pair aligned isomeric OMKLs $\mathcal{G} = ((G^{(1)}, G^{(2)}), (\mathcal{A}^{(1,2)}))$ as the example, and a simple extension of the proposed framework can also be applied to the case of multiple (more than two) aligned isomeric OMKLs as well.

\noindent \textbf{Missing Movie Identification Problem}: Given a pair of OMKLs $G^{(1)}$ and $G^{(2)}$, the \textit{missing movie identification} problem aims at uncover the movies which exist in one single movie knowledge library, i.e., the movies which exist in $G^{(1)}$ (or $G^{(2)}$) but don't exist in $G^{(2)}$ (or $G^{(1)}$). For simplicity, in this paper, we name the movies which exists $G^{(2)}$ but doesn't exist in $G^{(1)}$ as the \textit{missing movies} for $G^{(1)}$, and similarly we call the movies merely existing for $G^{(1)}$ as the \textit{missing movies} for $G^{(2)}$. 

In this paper, we propose to identify the missing movies for $G^{(1)}$ and $G^{(2)}$ simultaneously by inferring the potential \textit{movie anchor links} between $G^{(1)}$ and $G^{(2)}$. Let $\bar{\mathcal{A}}^{(i,j)}$ be the anchor link set (including both the known and inferred anchor links), the missing movies candidates for $G^{(1)}$ and $G^{(2)}$ can be represented as $\mathcal{M}^{(1)}_{m}$ and $\mathcal{M}^{(2)}_m$ respectively:\begingroup\makeatletter\def\f@size{7.5}\check@mathfonts
\begin{align*}
\mathcal{M}^{(1)}_{m} &= \{m^{(2)}_q | m^{(2)}_q \in \mathcal{M}^{(2)} \land (m^{(1)}_p, m^{(2)}_q) \notin \bar{\mathcal{A}}^{(1,2)}, \forall m^{(1)}_p \in \mathcal{M}^{(1)}\},\\
\mathcal{M}^{(2)}_m &= \{m^{(1)}_p | m^{(1)}_p \in \mathcal{M}^{(1)} \land (m^{(1)}_p, m^{(2)}_q) \notin \bar{\mathcal{A}}^{(1,2)}, \forall m^{(2)}_q \in \mathcal{M}^{(2)}\}.
\end{align*}\endgroup

\noindent \textbf{Missing Movie Ranking Problem}: Given the set of identified missing movies in $G^{(1)}$ or $G^{(2)}$, i.e., $\mathcal{M}^{(1)}_{m}$ and $\mathcal{M}^{(2)}_{m}$, the \textit{missing movie ranking problem} aims at ranking the missing movies in $\mathcal{M}^{(1)}_{m}$ and $\mathcal{M}^{(2)}_{m}$ according to certain ranking criteria. For the missing movies in $\mathcal{M}^{(1)}_{m}$ and $\mathcal{M}^{(2)}_{m}$, those achieving high ranking scores will be placed at the top of the recommendation list for completion.

Depending on the certain objective of the OMKLs, the ranking criteria can be quite diverse. For instance, to cater to the young people, the ranking criterion can be the movie genres/topics or actors/actress involved that the young people like the most. Meanwhile, from the movie artistic value or quality perspectives, the ranking criterion can be the moving rating received. If the OMKLs want to involve more popular movies, the ranking criterion can be the metrics like the review number in other OMKLs. On the other hand, in terms of maximizing the OMKLs' knowledge integrity, the movie ranking criterion will be the knowledge gap filling up the movies. In this paper, we will try two different two different ranking criteria in ranking the movies, \textit{movie popularity},  and \textit{movie quality}, which are quantitatively measured by \textit{movie comment number} and \textit{movie rating score} respectively.

\section{Online Movie Knowledge Library Statistical Analysis} \label{sec:general_analysis}

Before introducing the methods to solve the {\problem} problem, in this section, we will first study the Douban, IMDB datasets to provide some statistical analysis about these two \textit{multiple aligned isomeric OMKLs}. Analysis to be introduced about movies in Douban and IMDB are mainly focused on several important aspects, like the movie genres, review comments and review ratings in Douban and IMDB respectively, which will provide fundamental insights for building the {\our} framework to be introduced in the next section.

\subsection{Dataset Description}

\begin{table}[t]
\caption{Properties of the Different OMKLs.}
\label{tab:datastat}
\centering
\begin{tabular}{clrr}
\toprule
&&\multicolumn{2}{c}{OMKLs}\\
\cmidrule{3-4}
&property &\textbf{Douban}	&\textbf{IMDB}	 \\
\midrule 
\multirow{5}{*}{\# node}
&movie		& 14,829		& 14,485 		\\
&user		& 808,322		& 557,821 	\\
&comment	& 10,296,222	& 1,647,967 	\\
&genre		& 40			& 27			\\
&cast		& 25,065		& 377,965		\\
\midrule 
\multirow{6}{*}{\# link}
&movie-comment		& 10,296,222	& 1,647,967	\\
&user-comment		& 10,296,222	& 1,647,967	\\
&movie-genre			& 34,882		& 32,820		\\
&movie-director			& 9,911		& 31,015 		\\
&movie-writer			& 13,072		& 54,819 		\\
&movie-actor/actress	& 60,286		& 976,237		\\
\midrule 
\multirow{2}{*}{\# ancor link}
&movie anchor link		& 14,485	& 14,485	\\
&cast anchor link		& 25,065	& 25,065	\\
\bottomrule
\end{tabular}
\end{table}

\begin{figure*}
	\centering
	\includegraphics[width=1.0\textwidth]{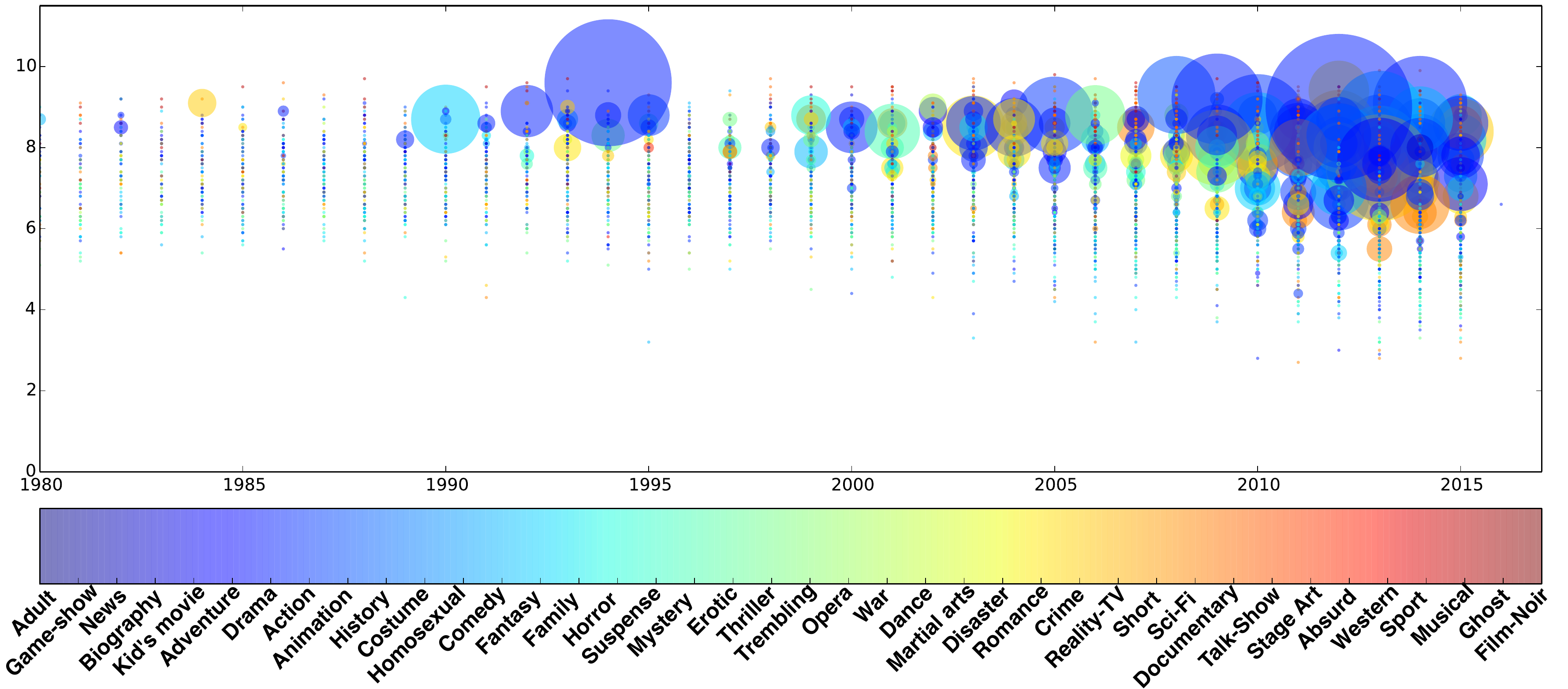}\vspace{-10pt}
	\caption{Douban Information Distribution about Year, Rating, Comment Number, and Genre.}\vspace{-10pt}
	\label{fig:douban_temporal}
\end{figure*}

The datasets used in this paper include two OMKLs: Douban and IMDB. Some basic statistical information about the datasets is available in Table~\ref{tab:datastat}. 

\begin{itemize}

\item \textit{Douban}: Currently, \textit{Douban} is the largest OMKL in China. Douban systematically organize existing and on-show movies into different categories according to different categorization criteria, like movie genre, production country, production year and even the main actors involved. Currently, Douban involves more than $153,000$ movies in all and the number is still keeping increasing in recent years. In this paper, to compare Douban and IMDB together with the audiences' opinions towards the movies in China and US, we crawled the movies which have been on show (or have released CDs for sale)  in both US and China, whose number is $14,829$. About these $14,829$ movies, $10,296,222$ comments have been posted by $808,322$ users, and each movie in Douban has received $694$ review comments on average. Fom Douban, $25,065$ cast crew members have been identified, and specific roles of the cast members in different movies can be \textit{director}, \textit{writer} and \textit{actor}/\textit{actress}, whose corresponding numbers are $9,911$, $13,072$ and $60,286$. These $14,829$ crawled movies belong to $40$ different genres in Douban, and each movie can be associated to multiple genres. The total number of movie-genre pair in the Douban dataset is $34,882$.

\item \textit{IMDB}: Among the $14,829$ crawled Douban movies, $14,485$ of them have their corresponding home pages in IMDB, from which we can obtain the general information as well. These $14,485$ movies in IMDB have received $1,647,967$ review comments from the $557,821$ IMDB users in all. On average, each movie obtains more than $113$ review comments in IMDB, and each user has posted about $3$ comments. In IMDB, the cast crew member information is more complete. For these $14,485$ IMDB movies, $377,965$ casts have been identified. Similarly, the roles of these identified cast members can be director, writer and actor/actress in these movies, whose corresponding pair numbers are $31,015$, $54,819$ and $976,237$ respectively. Among these $377,965$, $25,065$ also have their corresponding cast personal home pages in Douban, and the correspondence relationships between the identified casts in Douban and IMDB is denoted as the \textit{cast anchor links} (with \textit{one-to-one} cardinality constraint) in the dataset, whose number is $25,065$. Movies in IMDB are divided into $27$ different movie genres, and the total number of movie-genre pair numbers in IMDB is $32,820$. 

\end{itemize}

\subsection{General Movie Information Statistics}

In this section, we will provide the general analysis about the Douban Movie and IMDB Movie datasets in detail. The data distributions of these two movie datasets will be studied from several aspects of the movies: production year, movie genre, review comments and review rating. We will also compare the differences between the Chinese and the US movie audience/reviewers regarding the movies, which can provide fundamental insights for us in studying the movie completion problem.

The film industry is undergoing a vigorous development in recent years, especially in developing countries like China. New movies are keeping appearing in every year, some of which have achieved great reputations and remarkable revenues in terms of box office worldwide. In this section, we will study the general ratings and received review comments received from the Chinese and US audiences regarding different genres of movies produced in recent years (ranging from 1980 to 2016). 

\begin{figure*}
	\centering
	\includegraphics[width=1.0\textwidth]{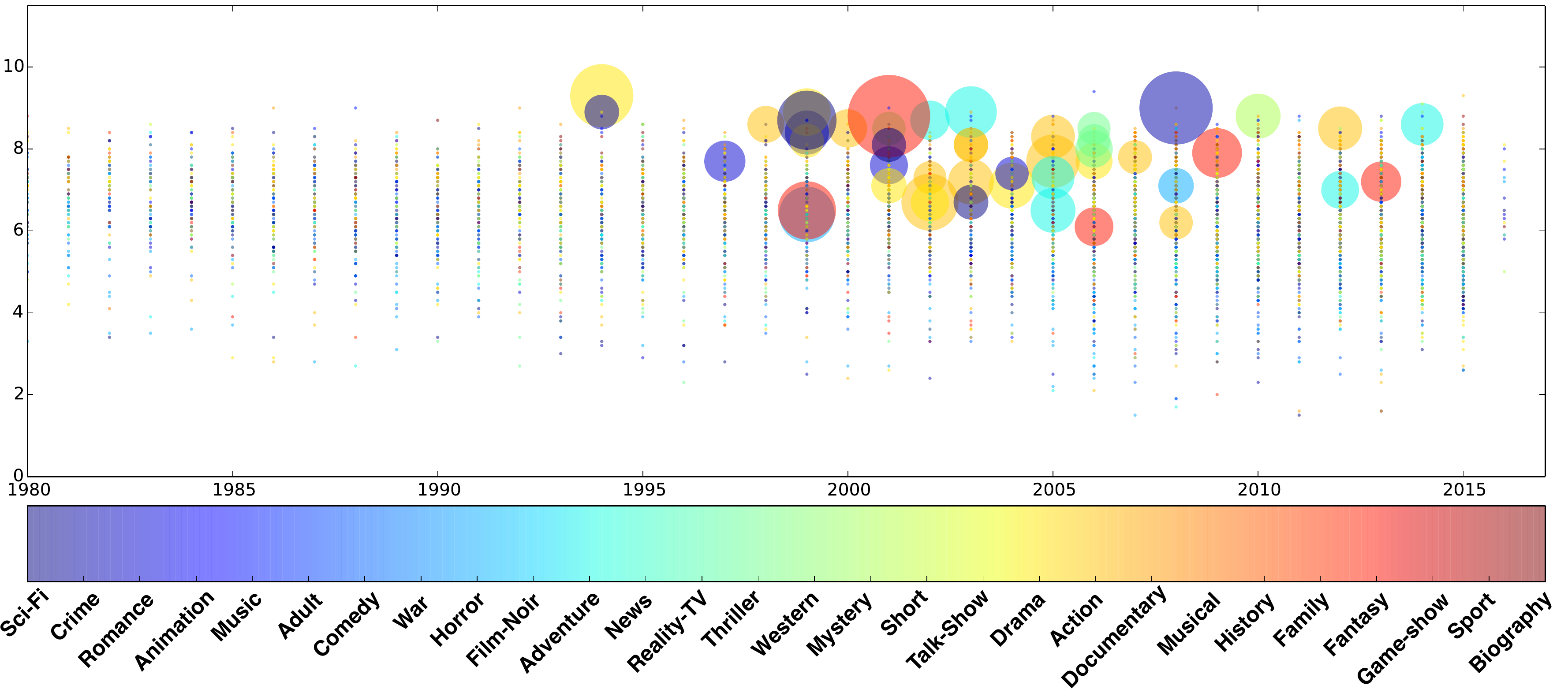}\vspace{-10pt}
	\caption{IMDB Movie Information Distribution about Year, Rating, Comment Number, and Genre.}\vspace{-10pt}
	\label{fig:imdb_temporal}
\end{figure*}

The results are shown in Figure~\ref{fig:douban_temporal} and Figure~\ref{fig:imdb_temporal} respectively. In the figures, we provide the information distribution of the same set of movies in Douban and IMDB in terms of their production years, overall rating scores and comment numbers. To different movies of different genres (like action movies, fiction movies and comedy movies), we use different colors in the figure and the correspondence relationship between movie genres and their colors is shown in the color bar at the bottom of the figures. In the plot, each circle denotes a movie, whose x axis and y axis coordinates denote the production year and rating of the movie. Meanwhile, the circle diameters represent the numbers of review comments posted by the Chinese/US audiences in Douban/IMDB about the movies (larger circles correspond to movies with more review comments).

\subsubsection{Information Analysis of Douban Movies and Chinese Audiences}

According to Figure~\ref{fig:douban_temporal}, We observe that the number of movies produced in recent years are keeping increasing. For instance, according to our datasets, the numbers of movies produced in years 1980, 1990, 2000, 2010, and 2015 are $49$, $115$, $210$, $315$ and $365$ respectively. Besides the movie numbers, we also have several important observations regarding the movie review comment, rating and genre information.

\vspace{3pt}
\noindent \textbf{Review Comments Focus on Recent Movies}: According to the movie comment data, we observe that the review comments are mainly about the recent movies (i.e., the circles in recent years are much larger). According to the plot, we count the number of review comments posted for movies produced in years 1980, 1990, 2000, 2010, and 2015 are $13,963$, $73,056$, $15,4301$, $665,562$ and $501,512$ respectively. Especially, for the movies produced in 2015, they can accumulate far more review comments than those produced before 2010 within merely one year. In addition, the review comments posted for movies produced in recent 10 years account for more than 70\% of the total review comments. Among the top $10$ movies receiving the most review comments, 7 of them are produced within the recent 5 years. Movie ``\textit{Life of Pi}'' achieves the most review comments (\#: 90,249) followed by ``\textit{The Shawshank Redemption}'' (\#: 78,314), ``\textit{Finding Mr. Right}'' (\#: 66,026), ``\textit{Avatar}'' (\#: 60,604) and ``\textit{Interstellar}'' (\#: 58,065), which are all movies produced in recent 5 years except ``\textit{The Shawshank Redemption}'' (produced in 1994).

\vspace{3pt}
\noindent \textbf{Recent Movie Quality Varies Significantly}: According to the movie rating data, for the movies produced before 2000, majority of them can achieve the rating scores greater than $5$ (mainly between $5$ and $9$). Meanwhile, in recent years, lots of the movies can even get rating scores lower than $3$. To demonstrate our statement, we calculate the average ratings together with the standard deviation of the rating scores achieved by movies produced in years 1980, 1990, 2000, 2010 and 2015 respectively, whose results are $7.20 \pm 0.937$, $7.39 \pm 0.790$, $7.35 \pm 0.846$, $7.16 \pm 1.038$ and $6.60 \pm 1.351$ respectively. Generally, movies produced during 1990-2000 can achieve relatively higher rating scores with smaller deviations. Meanwhile, the rating scores of recent movies are of lower rating scores and the rating deviation grows up a lot. Among all the movies, the top $5$ movies achieving the highest rating scores are the documentary movies ``\textit{Mind Meld: Secrets Behind the Voyage of a Lifetime}'' (rating: 9.9), ``\textit{The Unanswered Question: Six Talks at Harvard}'' (rating: 9.9) and ``\textit{Alicia Keys: MTV Unplugged}'' (rating: 9.8), ``\textit{Michael Jackson: HIStory on Film - Volume II}'' (rating: 9.7) and ``\textit{Guns N' Roses: Live at the Ritz}'' (rating: 9.7), which are mostly classic documentary or musical movies produced more than more than $30$ years ago. On the other hand, the top $5$ movies receiving the lowest rating scores in Douban are ``\textit{Aliens vs. Avatars}'' (rating: 2.7), ``\textit{Battle for Skyark}'' (rating: 2.8), ``\textit{Forgive Me for Raping You}'' (rating: 2.8), ``\textit{100 Degrees Below Zero}'' (rating: 2.8) and ``\textit{Orc Wars}'' (rating: 2.9), which are all lousy Sci-Fi movies produced in recent 6 years.

\vspace{3pt}
\noindent \textbf{Movie Genre Distribution}: Generally, each movie can belong to multiple movie genres. In the Douban dataset, for all the movies, the top $5$ movie genres with the most movies include ``Drama'', ``Comedy'', ``Horror'', ``Love'', and ``Action''. Movies belonging to these $5$ movie genres account for more than $81\%$ of the total movies. Meanwhile, from Figure~\ref{fig:douban_temporal}, we observe that the movies achieving the highest scores in each year (i.e., the circles at the top) are in orange/red colors. By studying the dataset, in terms of rating scores, the movie genres achieve the highest average rating scores from the users are ``Talk Show'', ``Documentry'', ``Stage Art'', ``Musical Show'' and ``Reality Show'', whose average rating scores are all above $8.0$. In terms of the comment number (i.e., the diameter of the circles), we observe that the largets circles are mostly in blue color. According to the dataset, the top $5$ movie genres obtaining the most review comments are ``Disaster'', ``Adventure'', ``Sci-fi'', ``Children's Movie'', and ``Fantasy''. On average, movies belong to these $5$ genres can obtain more than $2,000$ review comment in Douban.

\begin{figure*}
	\centering
	\includegraphics[width=1.0\textwidth]{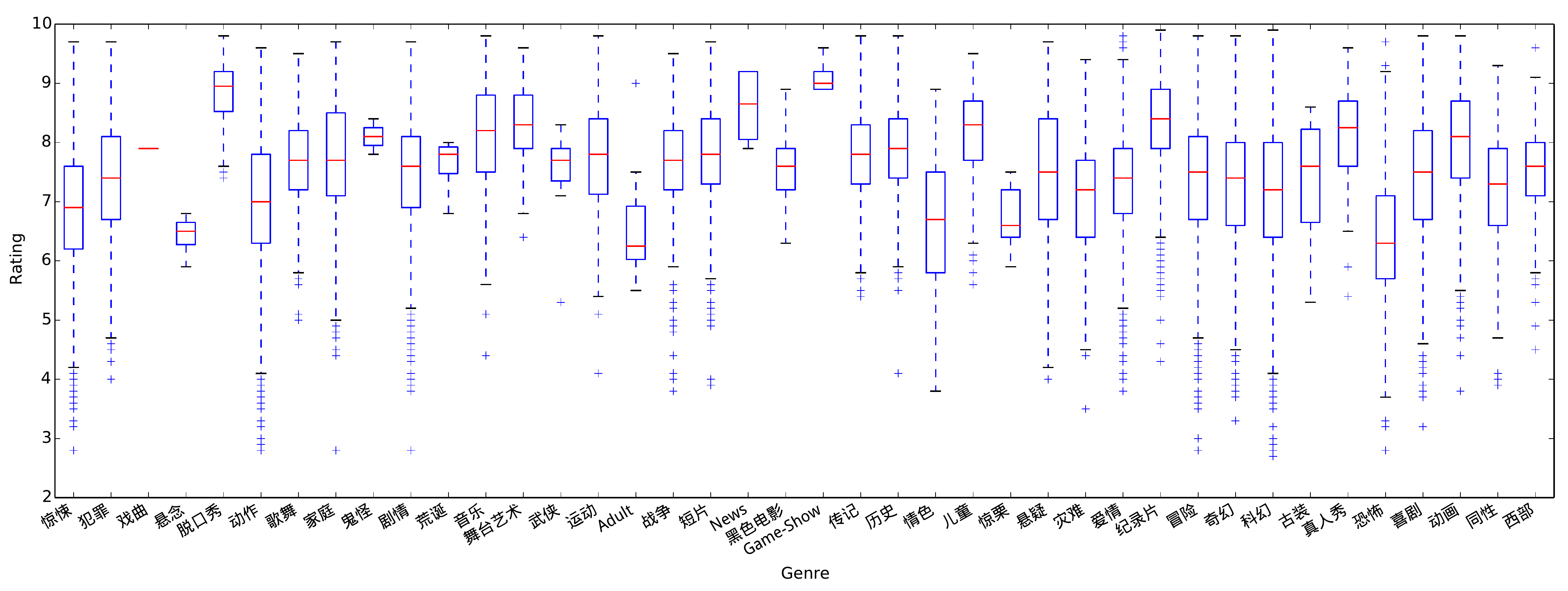}\vspace{-10pt}
	\caption{Rating Distribution of Douban Movie Genres.}\vspace{-10pt}
	\label{fig:douban_rating_genre}
\end{figure*}
\begin{figure*}
	\centering
	\includegraphics[width=1.0\textwidth]{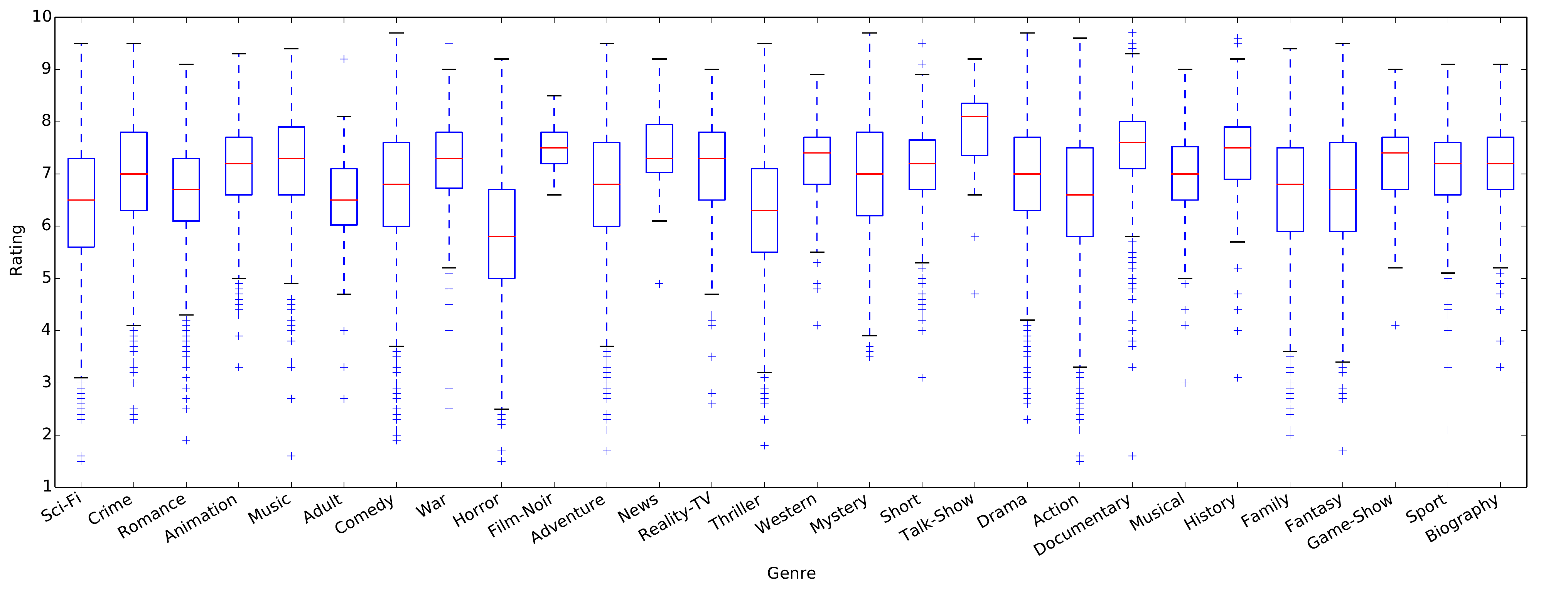}\vspace{-10pt}
	\caption{Rating Distribution of IMDB Movie Genres.}\vspace{-10pt}
	\label{fig:imdb_rating_genre}
\end{figure*}

\subsubsection{Information Analysis of IMDB Movies and US Audiences}

Meanwhile, for the same set of movies, the US audiences' review and rating information distributions are slightly different in IMDB. The analysis results are shown in Figure~\ref{fig:imdb_temporal}, which shows the rating, review number of movies belonging to different genres in each year. The meanings of the circle size, color and x axis and y axis are identical to that of Figure~\ref{fig:douban_temporal}. These movies from IMDB are identical to those studied in Figure~\ref{fig:douban_temporal}, and the number of movies produced in each year are keeping increasing since 1980s. However, as to the audiences' attitudes and reactions (measured by the ratings and review comments), the information distribution in IMDB is quite different from that in Douban.

\vspace{3pt}
\noindent \textbf{Movies Produced During 1997-2007 Receive the Most Comments}: By counting the review comments posted for movie produced in each year, we observe that movies produced during 1997 and 2007 can actually receive the most review comments, whose average review numbers are all above 150. What's more, movie produced in 2001 can obtain the most review comments, whose average number is 183.32 per movie. Different from Chinese audiences who like the most recent movies, the US movie audiences seem to be relatively rational. The average number of review comments posted on movies produced in recent 5 years is lower than 130, which display the significant differences about the audiences in the Chinese and US audiences and markets regarding the new movies. Among all the movies, the top $5$ movies receiving the most review comments are ``\textit{The Lord of the Rings: The Fellowship of the Ring}'' (\#: $5,027$), ``\textit{Star Wars: Episode I - The Phantom Menace}'' (\#: $4,462$), ``\textit{The Shawshank Redemption}'' (\#: $3,849$), ``\textit{The Matrix}'' (\#: $3,609$) and ``\textit{The Dark Knight}'' (\#: $3,544$), which all movies produced at least $15$ years ago except ``\textit{The Dark Knight}'' (produced in 2008).

\vspace{3pt}
\noindent \textbf{Movie Rating Doesn't Change Much In Different Years}: The movies in the US market have a significant distinction in terms of movie ratings before and after 1980. By studying the rating records of movies in the IMDB movie dataset, we observe that movies produced before 1980 (not shown in the figure) generally have a high rating score, which are all around 7.0 on average. For instance, the average rating scores of movies produced in 1979 is $6.95 \pm 0.882$, and that of movies produced in 1969, 1959 are $6.94 \pm 0.859$ and $7.03 \pm 1.333$ respectively.  Meanwhile, the movies produced after 1980 are of much lower rating scores, which stabilize around $6.5$ according to our records. For instance, the average rating scores of movies produced in 1980, 1990, 2000 and 2010 are at $6.69 \pm 1.028$, $6.38 \pm 0.971$, $6.46 \pm 1.059$ and $6.39 \pm 1.101$ respectively. Among all the movies in IMDB, the top $5$ movies achieving the highest rating scores are ``\textit{Yanni Live! The Concert Event}'' (rating: 9.4), ``\textit{The Shawshank Redemption}'' (rating: 9.3), ``\textit{Superhero Fight Club}'' (rating: 9.3), ``\textit{The Godfather}'' (rating: 9.2) and ``\textit{Shia LaBeouf Live}'' (rating: 9.1), two of which are actually just produced in the recent two years (\textit{Superhero Fight Club} in 2015 and \textit{Shia LaBeouf Live} in 2014). On the other hand, the top $5$ lousy movies achieving the lowest rating scores in our IMDB dataset are ``\textit{AVH: Alien vs. Hunter}'' (rating: 1.5), ``\textit{Aliens vs. Avatars}'' (rating: 1.5), ``\textit{Justin Bieber: Never Say Never}'' (rating: 1.6), ``\textit{From the Sea}'' (rating: 1.6) and ``\textit{Justin Bieber's Believe}'' (rating: 1.6), which are mostly Sci-Fi movies except two really bad music video about Justin Bieber.

\vspace{3pt}
\noindent \textbf{Movie Genre Distribution}: In IMDB, the movie genres are categorized in a totally different way as those in Douban. According to the dataset, IMDB divides movies into $27$ different genres, among which the top $5$ movie genres with the most movies are ``\textit{Drama}'', ``\textit{Comedy}'', ``\textit{Romance}'', ``\textit{Action}'' and ``\textit{Thriller}''. Movies belong to at least one of these $5$ genres account for more than $82\%$ of the total movies. Meanwhile, the top $5$ movie genres that contain movies of the highest rating scores include ``\textit{Talk-Show}'', ``\textit{Film-Noir}'', ``\textit{Documentary}'', ``\textit{News}'' and ``\textit{History}''. Movies belonging to the $5$ movie genres can generally achieve at least $7.2$ rating scores. Differently, the top $5$ movie genres that can get the most review comments from the US audiences are ``\textit{Sci-Fi}'', ``\textit{Adventure}'', ``\textit{Action}'', ``\textit{Fantasy}'' and ``\textit{Mystery}''. Movies of these $5$ genres can receive at least $200$ reviews in IMDB, while the average rating scores of movies in these $5$ genres are all lower than $6.5$.

\subsection{Douban and IMDB Difference Comparison with Further Statistical Analysis}

The general statistical analysis about the datasets have illustrate lots of interesting observations in the previous section. In this part, we will go further deep into the different movie genres, movie ratings and movie review comments to analyze the datasets and show the difference between Douban and IMDB more clearly.

\subsubsection{Difference in Movie Genres}

In Figures~\ref{fig:douban_rating_genre}-\ref{fig:imdb_rating_genre}, we show the movie rating scores of all the movie belonging to different genres, which can generally indicate the overall movie genre preference of audiences in China and US respectively. The results are presented with box plots, from which we can get a large amount of interesting information. In these two plots, the red bars in the boxes denote the median rating scores of each genre, the box sizes represent the interquartile ranges (IQR) covering the middle $50\%$ ratings, and the $+$ dots attached to the tails correspond to the outlier movies with scores exceeding $1.5 \times \mbox{IQR}$ ranges (i.e., the long tails).

\noindent \textbf{Douban Movie Genre Distribution}: From Figure~\ref{fig:douban_rating_genre}, by comparing the positions of the red line corresponding to each movie genre, we observe that the median rating scores fluctuate widely for different genres. For instance, the median rating scores of the \textit{game-show} and \textit{talk-show} movie genres are as high as $9.0$ and $8.95$ respectively, but those of the \textit{adult} and \textit{horror} movie genre are as low as $6.25$ and $6.3$. It indicate the rating scores is highly correlated to the topics of the movies, and those can make audiences happy can generally achieve higher rating scores (compared with those depressed ones). Even though the significant fluctuations, among all the $40$ movie genres, $60\%$ of them can achieve the median rating scores within $[7.0, 8.0)$, $22.5\%$ of the movie genres can achieve median rating scores within $[8.0, 9.0)$.

In the box plot, the box indicate the mid $50\%$ (i.e., the interquartile range (IQR) from the first quartile to the third quartile ) of the movie ratings belong to each genre. From the plot, we observe that, the box height of some movie genres, like \textit{opera}, \textit{suspense}, \textit{ghost} and \textit{game-show}, are relative short compared with the remaining movie genres. By studying the data, we observe that these movie genres are of a relatively small minority, and less than $10$ movies in total belong to these genres in our Douban dataset. 

Furthermore, from the plot, we observe that more outlier points are attached to the movie genres like \textit{action}, \textit{drama}, \textit{adventure} and \textit{sci-fi}, and the box tail of these movie genres are also relative much longer than other genres at the bottom. By analyzing the Douban dataset, we observe that more than $1,000$ movies are of these genres, and lots of the movies of these genres are really bad with very low rating scores. Meanwhile, from the plot, we observe that, for the genres like \textit{love}, \textit{adult}, \textit{horror} and \textit{western}, we also observe some outliers at the top, which indicate that there are some unexpectedly good movies of these genres achieving very high rating scores.

\noindent \textbf{IMDB Movie Genre Distribution}: In Figure~\ref{fig:imdb_rating_genre}, we show the box plot about the rating scores of the $27$ movie genres in the IMDB dataset. Compared with Figure~\ref{fig:douban_rating_genre}, there are several significant differences in the rating score distribution in terms of the median rating, the IQR range, and the outlier dots. In Figure~\ref{fig:imdb_rating_genre}, we observe that the median rating scores given to each movie genre in IMDB are relative lower than those in Douban. For instance, according to the distribution, the highest media rating score is $8.1$ achieved by the \textit{talk-show} movie genre, while the lowest median rating score $5.8$ is obtained by the \textit{horror} movie genre. The median rating scores of the remaining genres are all within range $[6.3, 7.6)$. In addition, the positions of the red bars corresponding to different movie genres are almost at relatively similar heights. With some simple statistics, among all the movie genres, the median rating score of $66.7\%$ movie genres are within range $[7.0, 7.5]$, the rating gaps among which are very small and negligible. 

In addition, box corresponding to each movie genres are of similar sizes. According to the datasets, movies belong to each genre are of a close number, except some small genres like \textit{adult}, \textit{news} and \textit{talk-show}. It indicates that the movie genre categorization in IMDB is relatively more reasonable, which effectively divides movies into different types and each of them can cover almost equal numbers of movies. 

Slightly different from Douban, in IMDB, the outlier dots attached to the box tails of the \textit{romance}, \textit{comedy}, \textit{drama} and \textit{documentary} genres are far more than the remaining genres. It indicates that the audiences in the US can actually appreciate and distinguish the quality of movies belong to these genres more easily compared with the Chinese audiences.

\subsubsection{Difference in Movie Review Comments}

\begin{figure}[t]
\centering
\subfigure[Douban]{\label{power_law_douban}
    \begin{minipage}[l]{.45\columnwidth}
      \centering
      \includegraphics[width=\textwidth]{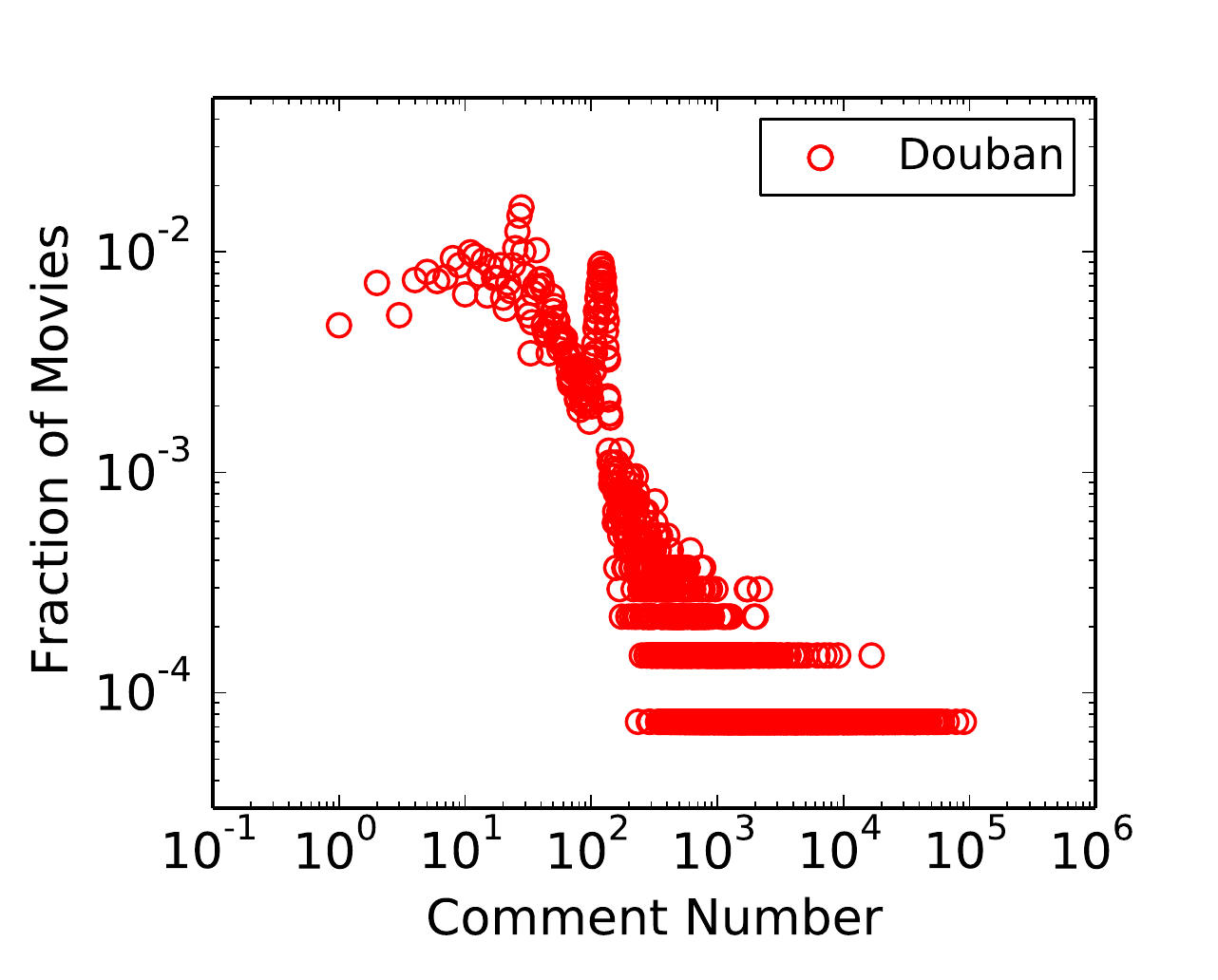}
    \end{minipage}
  }
  \subfigure[IMDB]{\label{power_law_imdb}
    \begin{minipage}[l]{.45\columnwidth}
      \centering
      \includegraphics[width=\textwidth]{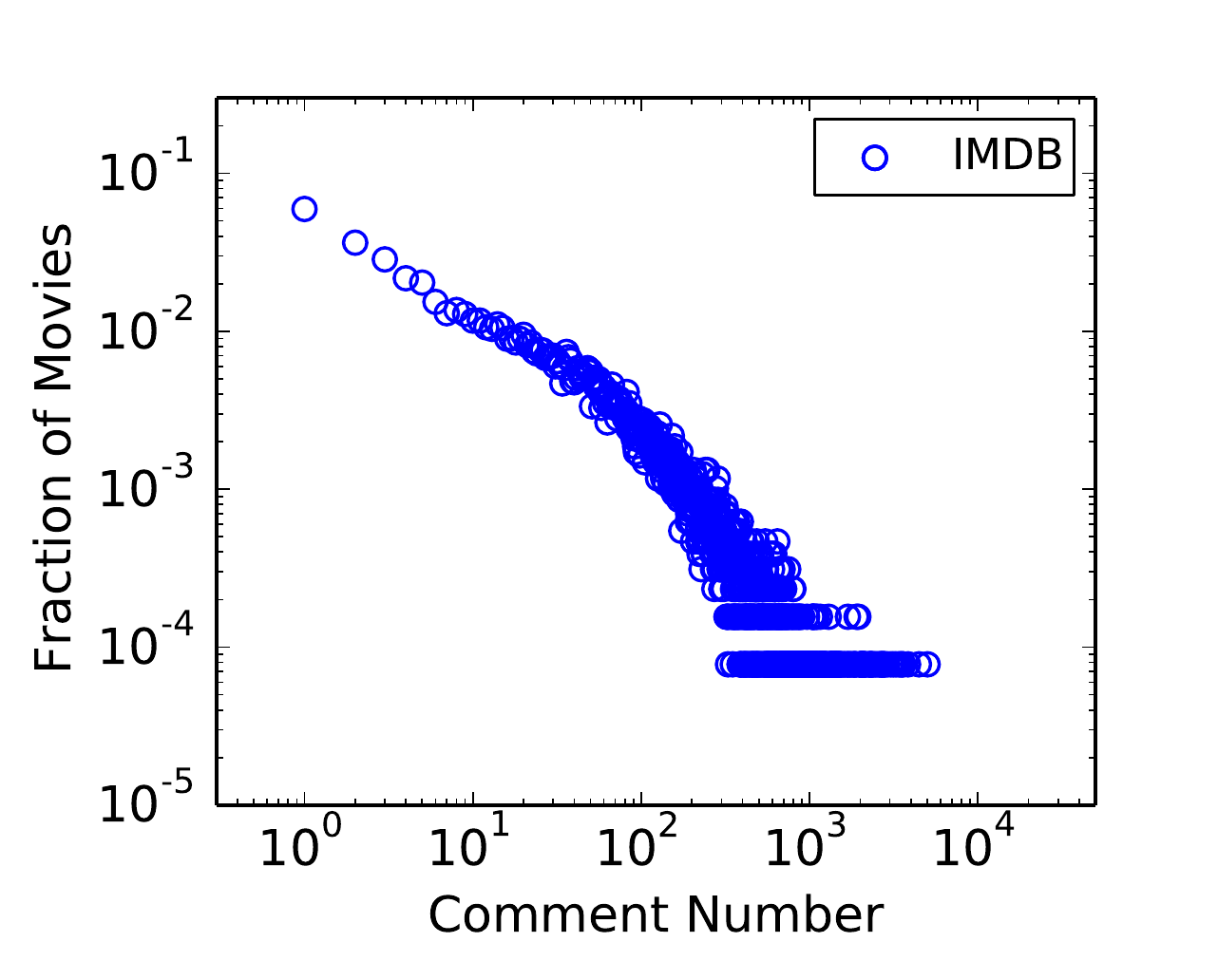}
    \end{minipage}
  }
\vspace{-5pt}
  \caption{Fraction of Movie vs Review Number Distribution of Movies in Douban and IMDB.}\label{fig:power_law}\vspace{-10pt}
\end{figure}

In this part, we will provide deeper analysis about the review comments posted for each movie in Douban and IMDB respectively. The analysis results about the fraction of movies achieve certain numbers of review comments in these two sites are available in Figures~\ref{power_law_douban}-\ref{power_law_imdb}. From the plots, we observe that the distribution plots of Douban and IMDB generally follow the power law distribution. Majority of the movies can receive less than $100$ review comments, but some of the movies can receive more than $50,000$ and $5,000$ comments in Douban and IMDB respectively. Generally, the review comments posted for movies in Douban and IMDB are not at the same level in terms the scale. The Chinese audiences seem to be more active at commenting on movies, and far more review comments have been posted in Douban.

\begin{figure}
	\centering
	\includegraphics[width=0.48\textwidth]{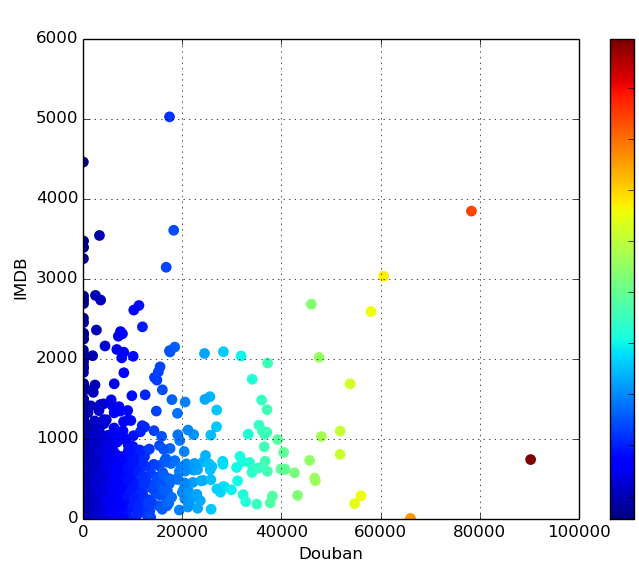}\vspace{-10pt}
	\caption{Review Number Comparison of Movies in Douban and IMDB}\vspace{-10pt}
	\label{fig:review_comparison}
\end{figure}

What's more, regarding the same movie, to compare the review comments received from Douban and IMDB, we further provide a plot in Figure~\ref{fig:review_comparison}. The x axis and y axis of the plot denote the review numbers received from Douban and IMDB respectively, and each dot in the plot represent a movie in the dataset. The color of the dot indicate how biased these movies are reviewed in Douban and IMDB (dark red color indicates that the movie is reviewed a lot in Douban but receives few comments in IMDB, and dark blue indicates the reverse). From the results, we observe that most of the dots are lied within the left bottom square, i.e., majority of the movies receive less than $20,000$ and $2,000$ review comments form Douban and IMDB. It is also in line with the results shown in Figures~\ref{power_law_douban}-\ref{power_law_imdb}.

To be more specific, the rightmost three dots in Figure~\ref{fig:review_comparison} denote to the movies receiving the most review comments in Douban, which are ``\textit{Life of Pi}'', ``\textit{The Shawshank Redemption}'' and ``\textit{Finding Mr. Right}'' respectively as introduced before. Meanwhile, the uppermost three dots in the plot denote the movies receiving the most review comments in IMDB instead, i.e., ``\textit{The Lord of the Rings: The Fellowship of the Ring}'' , ``\textit{Star Wars: Episode I - The Phantom Menace}'' and ``\textit{The Shawshank Redemption}''. Except ``\textit{The Shawshank Redemption}'' which is popular in both China and US, the comment posted for the other four movies are very biased, especially ``\textit{Finding Mr. Right}'' and ``\textit{Star Wars: Episode I - The Phantom Menace}''. It shows some very interesting differences in terms the movie popularity in these two different markets. By studying these two movies carefully, we notice that ``\textit{Finding Mr. Right}'' is actually a romantic Chinese movie, which is also on show in the US for a short period of time. This movie as well as the main actor and actress are all very well-known in China. However, the storyline of the movie and the cast are not what that US audiences prefer actually. As to the ``\textit{Star Wars: Episode I - The Phantom Menace}'', it is the first movie of the ``\textit{Star Wars}'' series and achieves remarkable success in the history of films in the US. However, this movie was on show in 1999, which is relatively too old for the Chinese audiences as well as Douban users.

\subsubsection{Differences in Movie Rating}

\begin{figure}
	\centering
	\includegraphics[width=0.48\textwidth]{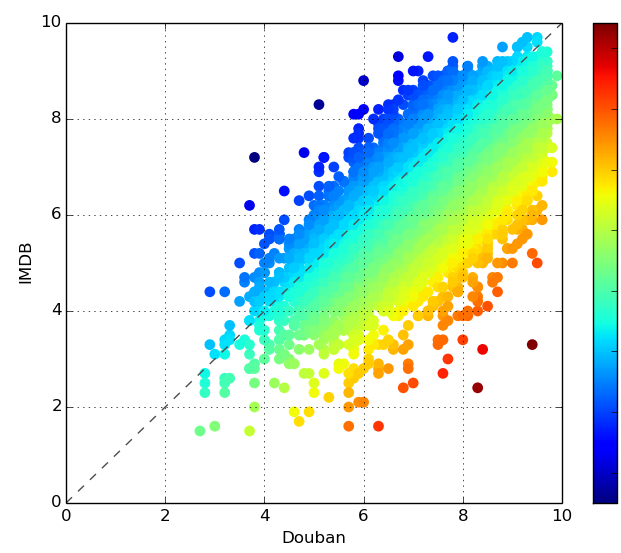}\vspace{-10pt}
	\caption{Rating Comparison of Movies in Douban and IMDB}\vspace{-10pt}
	\label{fig:rating_comparison}
\end{figure}

As to the audiences' preference about the movies in China and the US, we also have a deep study about the audiences' rating behaviors in these two countries about the same movies. The analysis results are shown in Figure~\ref{fig:rating_comparison}, where each dot represent a movie and its coordinates denote the rating score received in Douban and IMDB respectively. Here, the color of the dots indicate how biased these movies are rated in Douban and IMDB. The dark red color means that the movies with high ratings in Douban and low ratings in IMDB, while the dark blue color denotes the reverse. 

According to the analysis results, about $72.61\%$ of the movies can achieve rating scores with at most $1.0$ difference from Douban and IMDB respectively, which correspond to the movie dots close to the diagonal line of the plot. The Chinese audiences tend to give movies higher rating scores compared with the US audiences. From Figure~\ref{fig:rating_comparison}, we can observe that majority of the dots lie below the diagonal of the plot, i.e., majority of the movies can obtain larger rating scores in Douban. According to the data, about $78.4\%$ of the movies can achieve higher rating scores from Douban than that obtained from IMDB.

However, the rating scores of some of the movies can be quite biased. The top $3$ movie dots that are furthest away from the diagonal line at the lower-right corner (i.e., those with high ratings in Douban but low ratings in IMDB) are ``\textit{My Fair Madeline}'' (Douban: $9.4$, IMDB: $3.3$), ``\textit{Fright Club}'' (Douban: 8.3, IMDB: 2.4) and ``\textit{Inhumanoid}'' (Douban: 8.4, IMDB: 3.2). Animation movie ``\textit{My Fair Madeline}'' has been imported by China and broadcasted on TV many years ago, and many people give it a high rating due to their childhood memories. The other two are both horror movies about zombies. Movies of such a subject are relatively rare in China, and these two movies receive high rating scores maybe due to the audiences' curiosity in them. However, these three movies are rated with very low scores in the US. 

On the other hand, the top $3$ movies that are furthest away from the diagonal line at the upper-left corner (i.e., those with high rating in IMDB and low rating in Douban) are ``\textit{Kaptara}'' (Douban: $3.8$, IMDB: $7.2$), ``\textit{This Is How You Die}'' (Douban: $5.1$, IMDB: $8.3$) and ``\textit{Vishwaroopam}'' (Douban: $6.0$, IMDB: $8.8$). Adventure movie ``\textit{Kaptara}'' is a low-budget movie with an interesting storyline, which is to the taste of the US audience. However, the terrible special effects used in the movie make it unwelcome in the Chinese audience. Movie ``\textit{This Is How You Die}'' is of relatively strange and odd topic, which may seem to be interesting for the US audience but is quite boring for the Chinese audience.  ``\textit{Vishwaroopam}'' is actually an Indian movie and most of the high scores are rated by the Indian audiences in the US, which is actually not to the Chinese audiences' interest due to the cultural differences.

\subsubsection{Difference in Correlations between Movie Review Comment and Rating}

\begin{figure}[t]
\centering
\subfigure[Douban]{\label{result_2_1}
    \begin{minipage}[l]{.45\columnwidth}
      \centering
      \includegraphics[width=\textwidth]{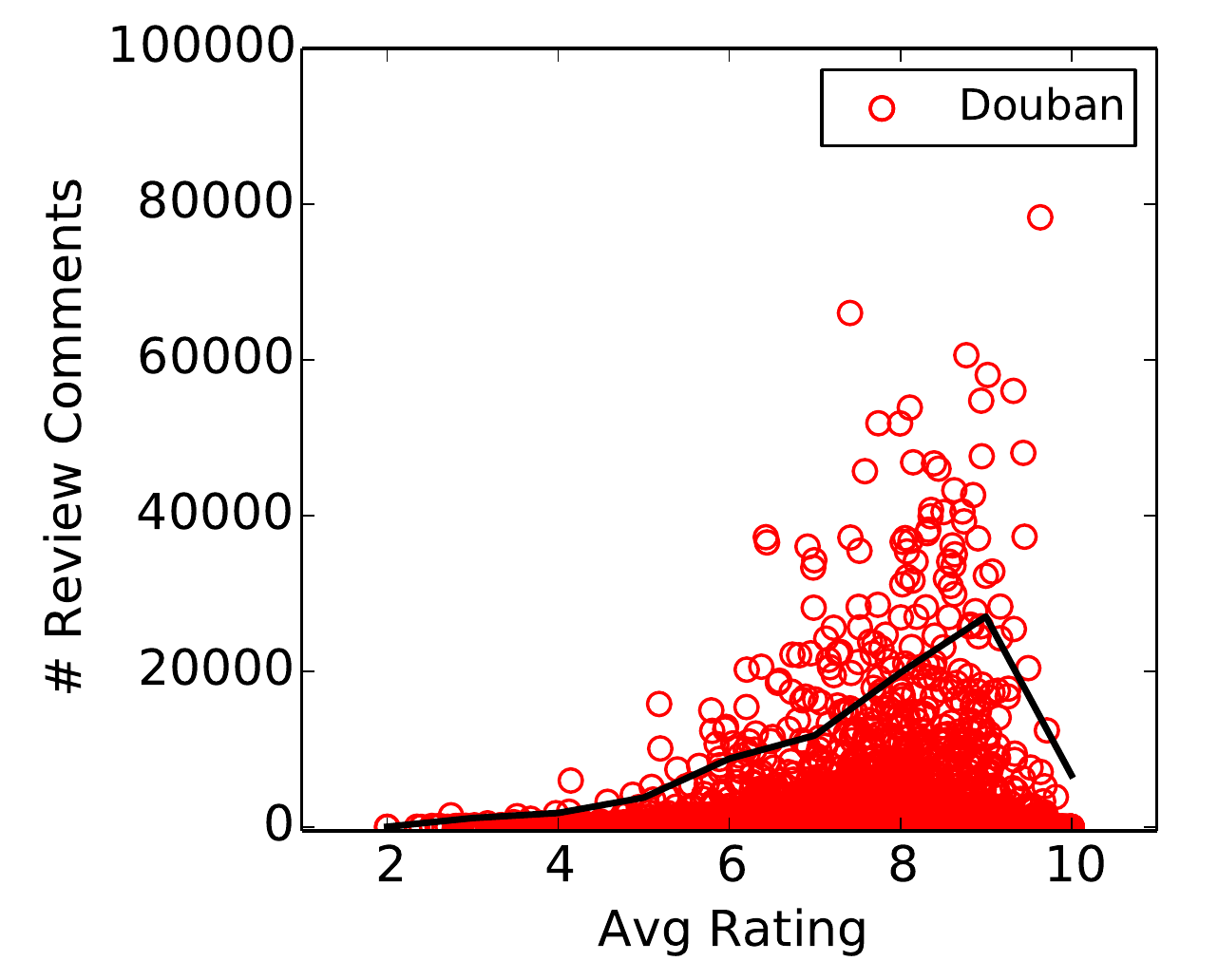}
    \end{minipage}
  }
  \subfigure[IMDB]{\label{result_2_2}
    \begin{minipage}[l]{.45\columnwidth}
      \centering
      \includegraphics[width=\textwidth]{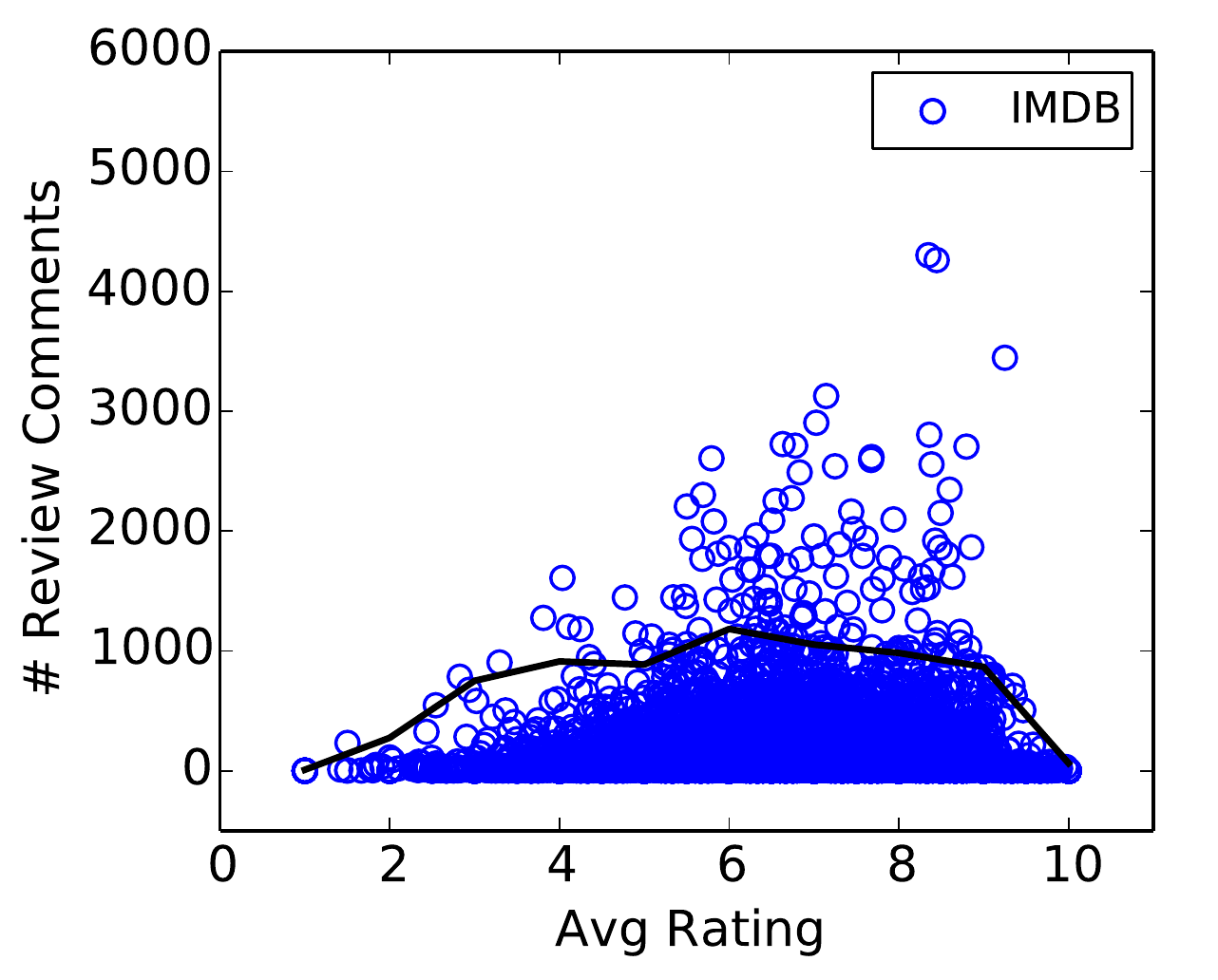}
    \end{minipage}
  }
\vspace{-5pt}
  \caption{Review Number vs Rating Distribution of Movies in Douban and IMDB.}\label{fig:review_number_rating}\vspace{-10pt}
\end{figure}

The users' review comments posted in Douban and IMDB effectively show the popularity of the movies in China and US. However, whether the review comment number can denote the quality of the movies or not is still an open questions. To study this problem, in this part, we will further study distribution of movies in terms of review number vs rating, and the results are shown in Figure~\ref{fig:review_number_rating}. In the plots, each circle represents a movie in the dataset, whose x and y axis coordinates are the number of review comments and rating score the movie receives from Douban and IMDB respectively. To show the distribution shape more clearly, a black poly lines covering $99\%$ of the movie instances for movies with ratings in $\{1.0, 2.0, \cdots, 10.0\}$ received from the users is also added to both of the plots in Figure~\ref{fig:review_number_rating}.

From Figure~\ref{result_2_1}, it is easy to observe that movies with high rating scores can generally obtain more review comments. For instance, on average, the comments received by movies with rating close to $8.0$ can be far more than the comments obtained by movies with rating score close to $9.0$. Generally, movies rated about $9.0$ can receive the most review comments. The reasons that the poly line goes down when x coordinate approaches $10.0$ is due to the fact that these highly rated movies (with scores close to $10.0$) are of minor genres, and few people actually watch them not to mention posting comments on them.

However, the comment rating plot about IMDB seems to be quite different from Douban. As shown in Figure~\ref{result_2_2}, the distribution is relatively flat for movies rated in range $[4.0, 9.0]$, and movies lie in this rating range can all get adequate review comments. In other words, the US movie audiences' comment posting behavior is not strongly correlated with their preference nor the quality about the movie. By studying the dataset, actually, for the movie with relative lower scores, people also like to post comments about them to express their opinions about the movies. In addition, in many case, the comments are about certain cast members, like the actor, actress or directors, and the US audiences are not that cared about the movie quality.



\subsection{Analysis Summary}

Based on the above analysis, we summarize our observations about the difference between Douban and IMDB and provide some analysis about the causes as follows:

\begin{itemize}

\item \textit{Movie Popularity}: As discovered in Figures~\ref{fig:review_comparison}, movies generally can receive more review comments in Douban compared with IMDB. From Figures~\ref{fig:douban_temporal}-\ref{fig:imdb_temporal}, we observe that Chinese audiences in Douban prefer recent movies the most, while the US audience in IMDB like the classic movies. The main reason causing such different observations can lie in the development of these two different markets. Over the few last decades, the rate of economic growth of the Chinese economy has been enormous compared to most, if not all, economies in developed countries. As a result, this growth has been evident in the film market (which keep growing by rate between $40\%$-$50\%$ between 2005-2011), where by 2012 China became the second largest consumer of feature films in the world in terms of box office. During the same period, the USA market has faced stagnation or experienced little increase in terms of admissions.

\item \textit{Movie Preference}: As discovered in Figure~\ref{fig:rating_comparison}, majority of the movies can achieve higher rating scores from the Chinese audience in Douban than the US audience from IMDB. In addition, from Figures~\ref{fig:douban_rating_genre}-\ref{fig:imdb_rating_genre}, we observe that the the Chinese audiences' preference about different movie genres fluctuate a lot, while the US audiences' preference about different movies genres tend to be more even. The reason causing such differences may come from the difference of the audiences in the China and US markets. Among the top $5$ movie genre of the highest scores in Figure~\ref{fig:douban_rating_genre} (i.e., ``\textit{Talk Show}'', ``\textit{Documentry}'', ``\textit{Stage Art}'', ``\textit{Musical Show}'' and ``\textit{Reality Show}''), 3 of them are the genres preferred by the young people. Meanwhile, for the top $5$ movie genres in Figure~\ref{fig:imdb_rating_genre}, (i.e., ``\textit{Talk-Show}'', ``\textit{Film-Noir}'', ``\textit{Documentary}'', ``\textit{News}'' and ``\textit{History}''), most of them are actually preferred by ``more senior people'' (instead of teenagers or young people). It also shows the difference between these two markets: China movie market is undergoing development and seems to be more energetic and prosperous involving more young people, while the US movie market is more mature and most movies are oriented at the adults.

\item \textit{Movie Preference vs Popularity}: As shown in Figure~\ref{fig:review_number_rating}, the correlation between the movie popularity (measured by comment number) and movie preference (measured by rating score) in the China and US markets are quite different. In both Douban and IMDB, the audiences don't like the movies of low quality and are reluctant to post review comments for them. Meanwhile, in Douban, the high quality movies are more popular among the Chinese audiences and can receive more review comments, especially for the movies with rating scores between $8$ and $9$. Different from the Chinese audience, the US movie audiences seem to be more mature. Generally, movies with rating scores between $4$-$9$ can all receive a sufficient amount of review comments from the US audiences. Such a difference may also due to the difference in the Chinese and US audience compositions. Generally, the young Chinese audiences tend to be more picky and like the good movies. When they are using Douban online, they are interested in differentiating these good movies from the remaining for the public via their reviews. On the other hand, the senior US audiences tend to be more inclusive and mature. They can appreciate the movies of different quality levels. When posting the comments online, they focus on more expressing their own feelings after watching the movies, even for those with mid-ranged rating socres.

\end{itemize}

\section{Missing Movie Identification and Ranking} \label{sec:method}

\begin{figure}
\vspace{-25pt}
	\centering
	\includegraphics[width=0.45\textwidth]{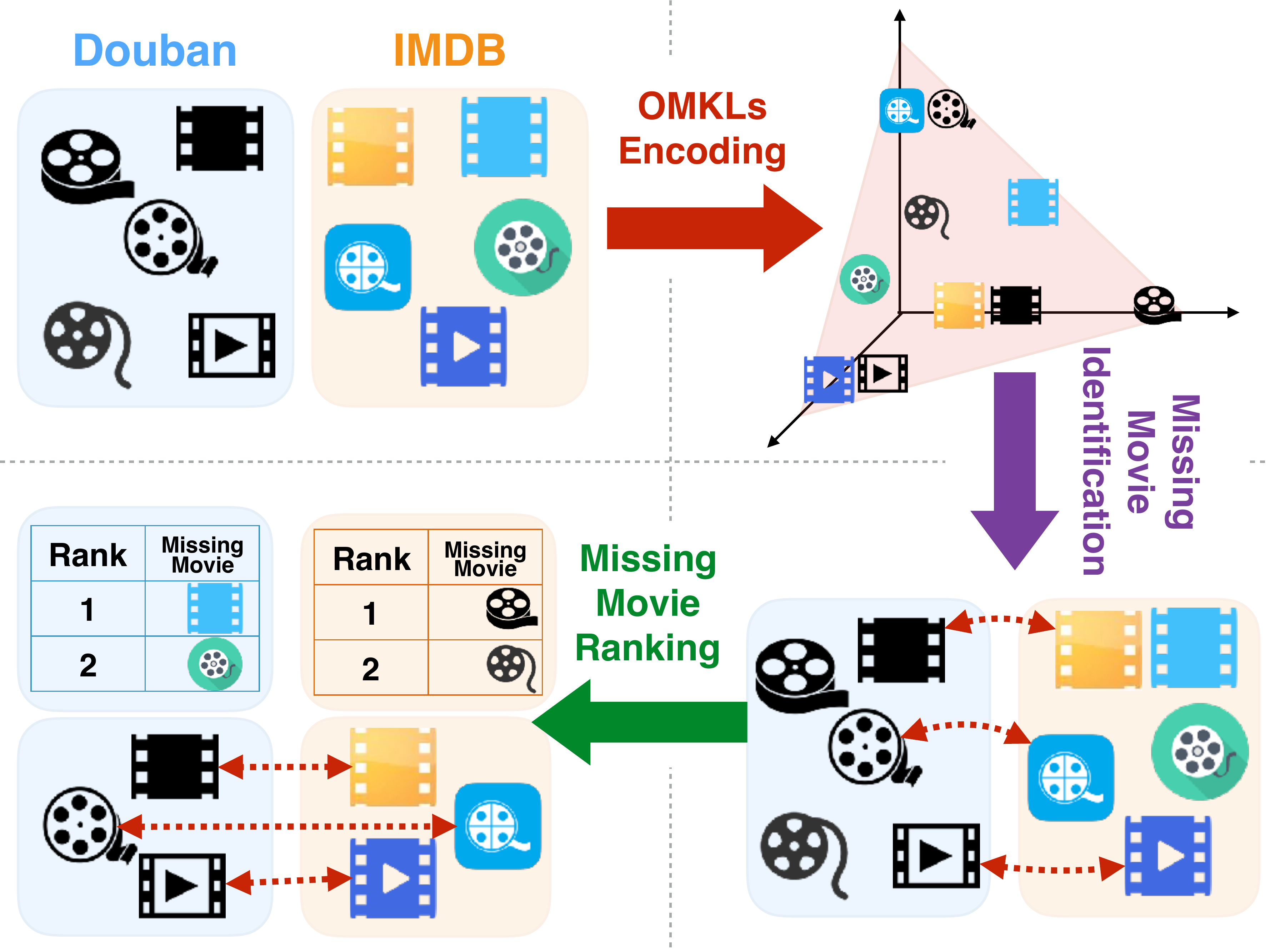}\vspace{-9pt}
	\caption{The Framework of {\our}.}\vspace{-15pt}
	\label{fig:framework}
\end{figure}

The statistical analyses provided in the previous section offers important signals for us to understand the Chinese and US audiences' preferences about different movies. Besides the shared interest in certain movies, the uniqueness of their preference should be taken into consideration when completing these OMKLs. To resolve the {\problem} problem, a novel missing movie identification and ranking framework {\our} will be introduced in this section. As shown in Figure~\ref{fig:framework}, framework {\our} involves three steps, including \textit{movie embedding}, \textit{movie fusion} and \textit{missing movie ranking}. In this part, we will introduce these three steps one by one in detail.

\begin{figure}
\vspace{-25pt}
	\centering
	\includegraphics[width=0.35\textwidth]{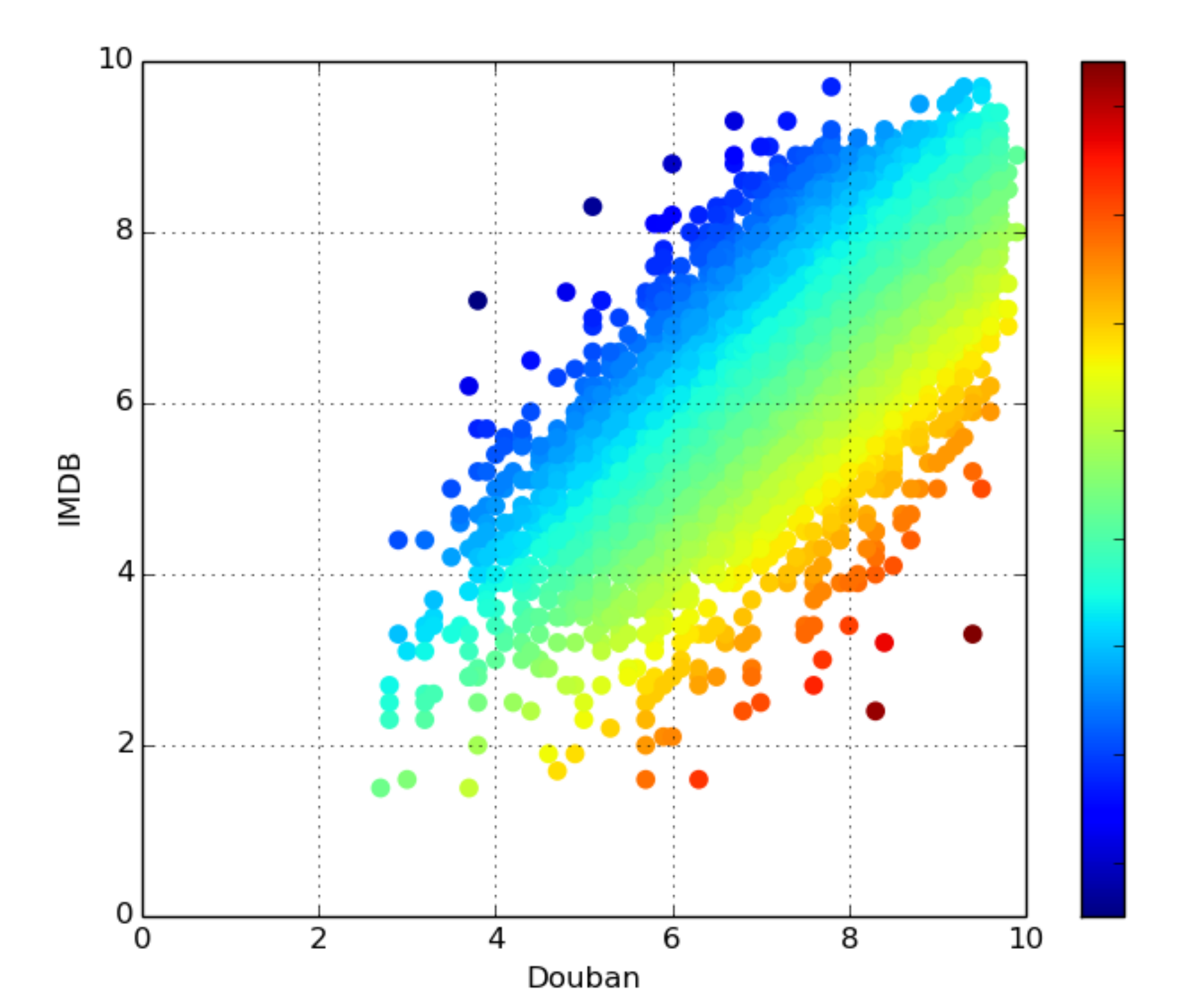}\vspace{-10pt}
	\caption{Rating Comparison of Movies in Douban and IMDB}\vspace{-15pt}
	\label{fig:rating_comparison}
\end{figure}


\subsection{Movie Embedding}\label{subsec:autoencoder}

For movies from different OMKLs, we can represent each of them as a feature vector, which describes the basic attributes of the movies as well as the review comments and rating scores received from the audience. The specific extracted features will be introduced in detail in Section~\ref{sec:experiment}. Since the information for the movies are collected from different isomeric sources, and the feature representation of each movie can be quite different. (For instance, the review comments in Douban are in Chinese, and the feature representation of the movie comments will be a bag-of-word in Chinese, which is totally different from the English bag-of-word feature representation of the same movie in IMDB.) Before carrying out the movie sources fusion and ranking, we propose to project the movies to a common feature space via embedding.

Formally, let $\{\mb{x}^{(1)}_i\}_{m^{(1)}_i \in \mathcal{M}^{(1)}}$ and $\{\mb{x}^{(2)}_j\}_{m^{(2)}_j \in \mathcal{M}^{(2)}}$ represent the set of input features extracted for movies in libraries $G^{(1)}$ and $G^{(2)}$ respectively ($\mb{x}^{(1)}_i \in \mathbb{R}^{d^{(1)}}$ and $\mb{x}^{(2)}_j \in \mathbb{R}^{d^{(2)}}$). Based on the input feature vectors $\mb{x}^{(1)}_i$ and $\mb{x}^{(2)}_j$, the final hidden representation of each movie (e.g., $m^{(1)}_i$ from $G^{(1)}$ or $m^{(2)}_j$ from $G^{(2)}$) in the deep autoencoder model can be represented as $\mb{y}^{(1), k}_i, \mb{y}^{(2), k}_j \in \mathbb{R}^{d}$, $d < d^{(1)}, d < d^{(2)}$ respectively, where $k$ denotes the number of layers in the encoding step. The encoded feature representations of movies from $G^{(1)}$ and $G^{(2)}$ are in the same feature space and of the same dimension. Furthermore, based on the input feature vectors $\mb{x}^{(1)}_i$ and $\mb{x}^{(2)}_j$, the representation of movies $m^{(1)}_i$ and $m^{(2)}_j$ in each hidden layer of the encoding step can be represented as
\begin{align*}
\begin{cases}
\mb{y}^{(1),1}_i &= \sigma(\mb{W}^{(1),1} \mb{x}^{(1),1}_i + \mb{b}^{(1),1}),\\
\mb{y}^{(1),m}_i &= \sigma(\mb{W}^{(1),m} \mb{y}^{(1),m-1}_i + \mb{b}^{(1),m}), m \in \{2, 3, \cdots, k\},
\end{cases}
\\
\begin{cases}
\mb{y}^{(2),1}_j &= \sigma(\mb{W}^{(2),1} \mb{x}^{(2),1}_j + \mb{b}^{(2),1}),\\
\mb{y}^{(2),m}_j &= \sigma(\mb{W}^{(2),m} \mb{y}^{(2),m-1}_j + \mb{b}^{(2),m}), m \in \{2, 3, \cdots, k\},
\end{cases}
\end{align*}
where matrices $\mb{W}^{(1), m}$ and $\mb{W}^{(2), m}$ denote the weights of connections among neurons in layers $k$ and $k+1$ and $\mb{b}^{(1), m}$ and $\mb{b}^{(2), m}$ are the bias terms. For simplicity, in the following parts, we will use notations $\mb{y}^{(2)}_j$ and $\mb{y}^{(1)}_j$ as the final representation of the encoded feature vectors $\mb{y}^{(2), k}_j$ and $\mb{y}^{(1), k}_j$ respectively.

After obtaining $\mb{y}^{(1)}_i$ and $\mb{y}^{(2)}_j$, we can reverse the calculation direction and start the decoder step and obtain the output $\hat{\mb{x}}^{(1)}_i$ and $\hat{\mb{x}}^{(2)}_j$, parameterized by $\hat{\mb{W}}^{(2),m}$, $\hat{\mb{W}}^{(2),m}$ and bias vectors $\hat{\mathbf{b}}^{(2),m}$ and $\hat{\mathbf{b}}^{(1),m}$ for $(m \in \{1, 2, \cdots k\})$. The goal of the decoder step is to minimize the reconstruction loss between the input feature vectors ${\mb{x}}^{(1)}_i$, ${\mb{x}}^{(2)}_j$ and the output feature vectors $\hat{\mb{x}}^{(1)}_i$, $\hat{\mb{x}}^{(2)}_j$. Formally, we can represent the loss function introduced by encoding movies in $\mathcal{M}^{(1)}$ and $\mathcal{M}^{(2)}$ via the deep autoencoder model as:\begingroup\makeatletter\def\f@size{7.5}\check@mathfonts
\begin{align*}
&\mathcal{L}_e (\mathcal{M}^{(1)}, \mathcal{M}^{(2)}) =\mathcal{L}_e(\mathcal{M}^{(1)}) + \mathcal{L}_e(\mathcal{M}^{(2)})\\
&= \sum_{m^{(1)}_i \in \mathcal{M}^{(1)}} \left\| (\hat{\mb{x}}^{(1)}_i - {\mb{x}}^{(1)}_i) \odot \mb{c}_i^{(1)} \right\|_2^2 +  \sum_{m^{(2)}_j \in \mathcal{M}^{(2)}} \left\| (\hat{\mb{x}}^{(2)}_j - {\mb{x}}^{(2)}_j) \odot \mb{c}_j^{(2)} \right\|_2^2.
\end{align*}\endgroup
In the above equation, vector $\mb{c}_i^{(1)} \in \{1, \gamma\}^{d^{(1)} \times 1}$ where $c_{i}^{(1)}(k) = 1$ if $x_{i}^{(1)}(k) = 0$; and $c_{i}^{(1)}(k) = \gamma$ ($\gamma > 1$) if $x_{i}^{(1)}(k) = 1$, which is added to avoid the trivial embedding results like $\mb{0}$ for the sparse inputs. By minimizing the loss function, we can learn the parameters in both the encoder and decoder steps.



\subsection{Missing Movie Identification}\label{subsec:fusion}

\begin{figure}
\vspace{-25pt}
	\centering
	\includegraphics[width=0.45\textwidth]{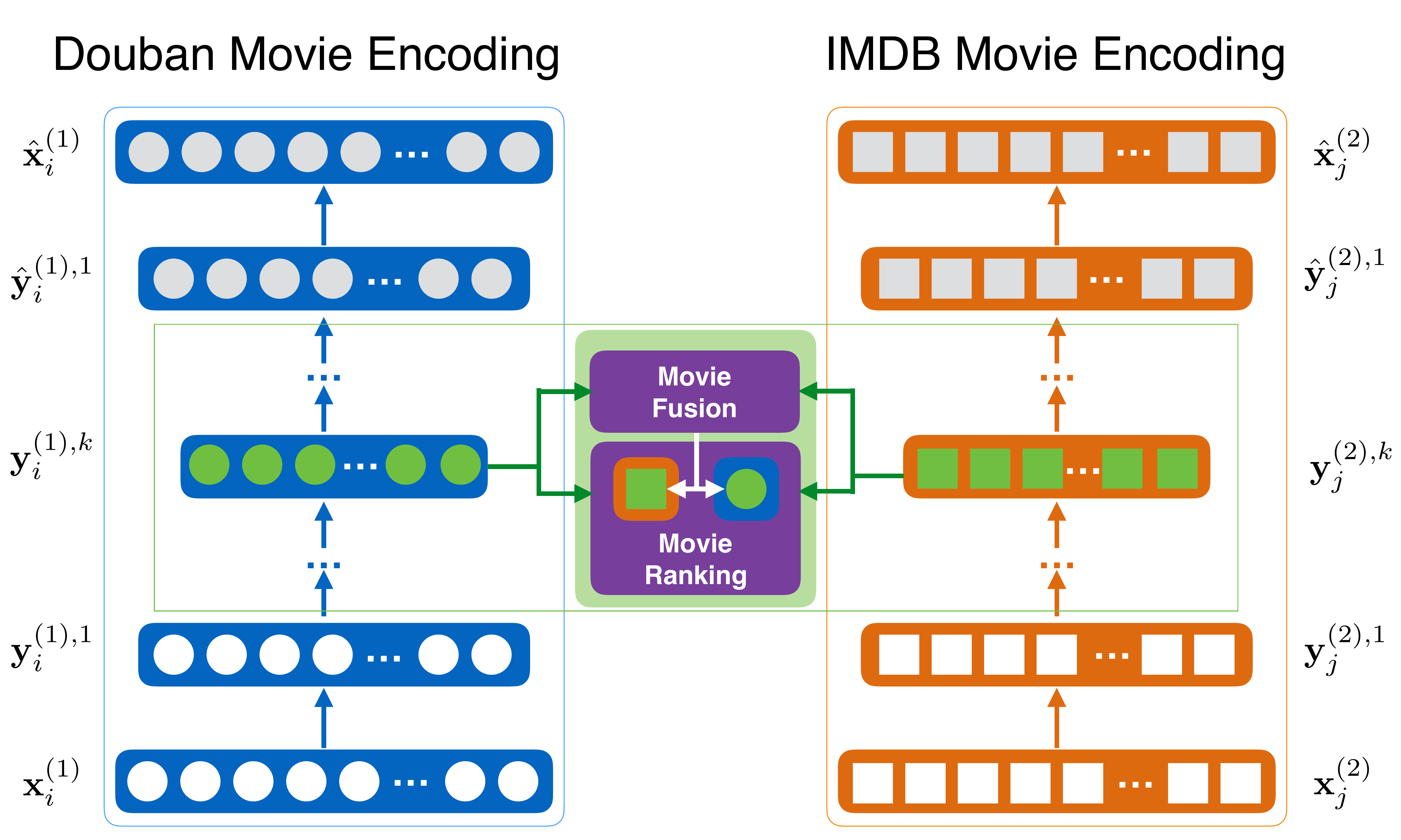}\vspace{-9pt}
	\caption{Deep Movie Fusion Model.}\vspace{-15pt}
	\label{fig:fusion}
\end{figure}

By projecting the movie information into a shared space via multiple non-linear mappings, the encoding vectors can capture the information representing the movie characteristics. In this part, we propose to fuse the multiple isomeric OMKLs by inferring the potential movie anchor links to identify the common movies existing in both $G^{(1)}$ and $G^{(2)}$ based on these encoding feature representations, which is achieved via model {\our}. In this subsection, we will introduce the {\our} model in detail, whose architecture is available in Figure~\ref{fig:fusion}.

Formally, given a training set involving the observed movie anchor links between OMKLs $G^{(1)}$ and $G^{(2)}$, $\mathcal{A} = \{(m^{(1)}_i, m^{(2)})_j\}$, we can represent the extracted feature vectors and binary labels for these pairs as tuples $\{ \left( (\mb{x}^{(1)}_i, \mb{x}^{(2)}_j ), s^{(1,2)}_{i,j} \right) \}_{(m^{(1)}_i, m^{(2)}_j) \in \mathcal{A}}$, where binary label $s^{(1,2)}_{i,j} = 1$ iff $m^{(1)}_i$ and $m^{(2)}_j$ are the same movie and $-1$ otherwise. For movies which are connected by the observed \textit{movie anchor links} in the training set, we aim at minimizing the difference of the inferred feature representation in the latent space. Formally, we can define the loss function introduced in aligning movies $m^{(1)}_i$ and $m^{(2)}_j$ from OMKLs $G^{(1)}$ and $G^{(2)}$ as
$$\mathcal{L}_f (m^{(1)}_i, m^{(2)}_j) = s^{(1,2)}_{i,j} \left\| \mb{y}^{(1)}_i - \mb{y}^{(2)}_j \right\|_2^2.$$

Furthermore, the loss function of aligning OMKLs $G^{(1)}$ and $G^{(2)}$ based on the observed training set $\mathcal{A}$ can be represented as
\begin{align*}
&\mathcal{L}_f (\mathcal{M}^{(1)}, \mathcal{M}^{(2)}; \mathcal{A}) = \sum_{(m^{(1)}_i, m^{(2)}_j) \in \mathcal{A}} \mathcal{L}_f (m^{(1)}_i, m^{(2)}_j) \\
&= \sum_{(m^{(1)}_i, m^{(2)}_j) \in \mathcal{A}} s^{(1,2)}_{i,j} \left\| \mb{y}^{(1)}_i - \mb{y}^{(2)}_j \right\|_2^2.
\end{align*}

Combined with the loss function in the embedding, we can represent the integrated loss function of the OMKLs fusion problem as follows:\begingroup\makeatletter\def\f@size{8}\check@mathfonts
\begin{align*}
&\mathcal{L} (\mathcal{M}^{(1)}, \mathcal{M}^{(2)}) \\
&= \mathcal{L}_e (\mathcal{M}^{(1)}, \mathcal{M}^{(2)}) + \alpha \cdot \mathcal{L}_f (\mathcal{M}^{(1)}, \mathcal{M}^{(2)}; \mathcal{A}) + \beta \cdot \mathcal{L}_{r} (\mathcal{M}^{(1)}, \mathcal{M}^{(2)}),
\end{align*}\endgroup
where parameters $\alpha$ and $\beta$ denotes the weight of the fusion loss and regularizer terms. Term $\mathcal{L}_r (\mathcal{M}^{(1)}, \mathcal{M}^{(2)})$ denotes the regularizer added to the function to avoid overfitting. In this paper, we apply Frobenius norm for the regularization and $\mathcal{L}_r (\mathcal{M}^{(1)}, \mathcal{M}^{(2)})$ can be represented as\begingroup\makeatletter\def\f@size{7.5}\check@mathfonts
\begin{align*}
&\mathcal{L}_{r} (\mathcal{M}^{(1)}, \mathcal{M}^{(2)}) = \mathcal{L}_{r} (\mathcal{M}^{(2)}) + \mathcal{L}_{r} (\mathcal{M}^{(2)}),\\
&= \sum_{m = 1}^{k} ( \left\| \mb{W}^{(1),m} \right\|_F^2 + \left\| \hat{\mb{W}}^{(1),m} \right\|_F^2 ) + \sum_{m = 1}^{k} ( \left\| \mb{W}^{(2),m} \right\|_F^2 + \left\| \hat{\mb{W}}^{(2),m} \right\|_F^2 ).
\end{align*}
\endgroup
To optimize the above objective function, we utilize Stochastic Gradient Descent (SGD). To be more specific, the training process involves multiple epochs. In each epoch, the training data is shuffled and a minibatch of the instances are sampled to update the parameters with SGD. Such a process continues until either convergence or the training epochs have been finished.

\subsection{Missing Movie Ranking}\label{subsec:ranking}
%

For all the missing movies identified in the previous subsection, some of them are not welcome in certain countries. For instance, movies which are identified to be politically incorrect (e.g., movie with racial discrimination in the US, and movies promoting national division in China), are highly likely to be deleted shortly after the creation. Therefore, creating of such missing movies in the knowledge libraries is a waste of time, and we should focus on movies which are considered to be useful, e.g., filling an important knowledge gap. Providing a useful ranking of these identified missing movies for creation can be of great importance for the service providers.

Depending on the specific settings and objectives, the ranking criterion can be very diverse. In this paper, two different ranking criterion, movie popularity and movie quality, are considered. We propose to measure the movie popularity and quality based on their review comment numbers and rating scores respectively. For instance, when using the movie popularity as the ranking objective, we introduce the concept of \textit{movie quality rank} in this paper and define it as follows:
$$r_{q}(m^{(1)}_i) = \frac{\mbox{rank}(m^{(1)}_i, \mbox{rating})}{|\mathcal{M}^{(1)}|},$$
where $\mbox{rank}(m^{(1)}_i, \mbox{rating})$ denotes the relative rank position of movie $m^{(1)}_i$ among all the movies in knowledge library $G^{(1)}$ in terms of rating scores.

\begin{figure}[t]
\vspace{-25pt}
\centering
\subfigure[AUC]{ \label{fig:anchor_link_1}
    \begin{minipage}[l]{0.45\columnwidth}
      \centering
      \includegraphics[width=1.0\textwidth]{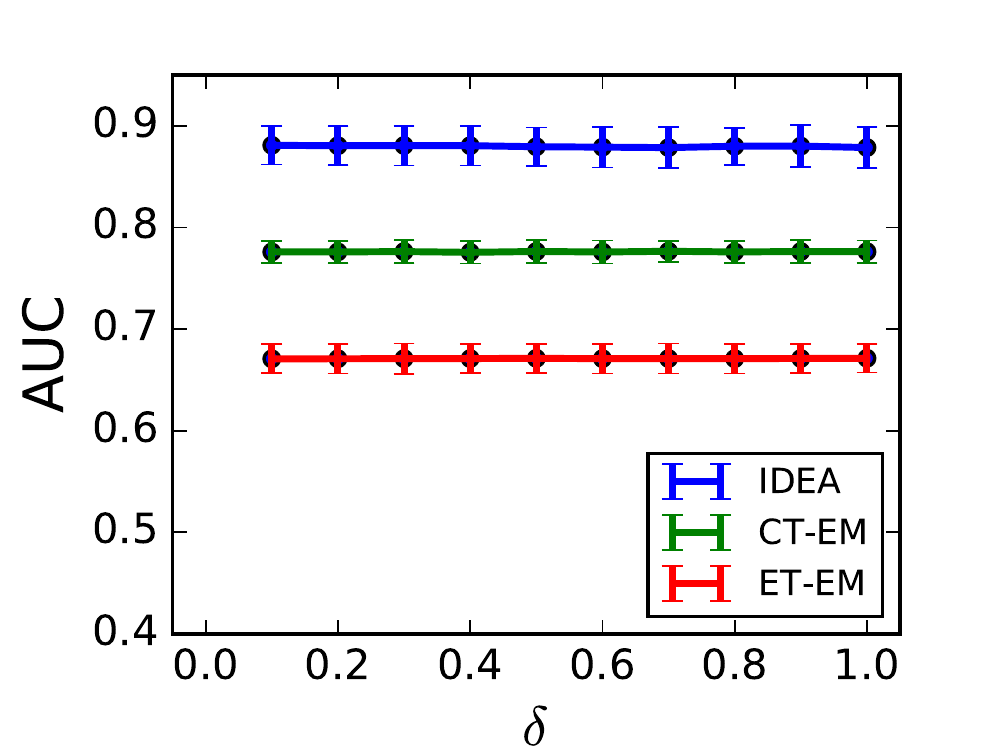}
    \end{minipage}
}
\subfigure[Prec@K]{ \label{fig:anchor_link_2}
    \begin{minipage}[l]{0.45\columnwidth}
      \centering
      \includegraphics[width=1.0\textwidth]{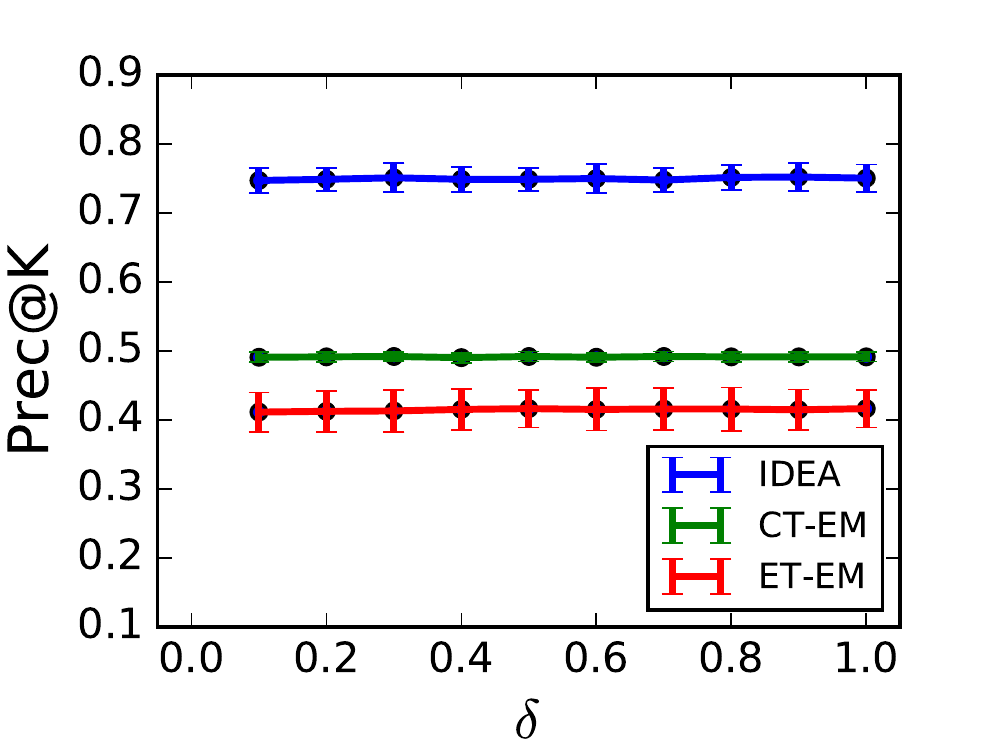}
    \end{minipage}
}\vspace{-5pt}
\caption{Anchor link prediction with various $\delta$.}\label{fig:anchor_link}\vspace{-15pt}
\end{figure}

\textit{Movie quality rank} is a normalized movie ranking criteria with values in range $[0, 1]$ and is applicable in multiple OMKLs, where high-quality movies will have ranking values close to $0$, while the low-quality movies will have ranking scores close to $1$ instead. In a similar way, we can define the \textit{movie popularity rank} concept based on the movie review comment numbers, which can be denoted as $r_{p}(m^{(1)}_i)$ for movie $m^{(1)}_i$. To be general, in the following parts, we will use term $r(m^{(1)}_i)$ to denote the rank of movie $m^{(1)}_i$, which can be either the quality or popularity based metric depending on the specific setting.

In this paper, we propose to build the model {\our} based on embedding features obtained from the first step. We aim at inferring the ranking scores of the movies based on their encoded features. As shown in Figure~\ref{fig:fusion}, when building the {\our} ranking models for Douban and IMDB OMKLs respectively, we need to swapping their input feature vectors and their ranking metrics. For instance, when training the {\our} model for Douban (i.e., $G^{(1)}$), for all the movies which exists in both Douban and IMDB (e.g., $m$, which corresponds to $m^{(1)}_i$ in Douban and $m^{(2)}_j$ in IMDB), we use its embedding feature vector from IMDB (i.e., $\mb{y}^{(2)}_j$) while its rank from Douban (i.e., $r^{(1)}_i$) to build the {\our} model. After building the model, if we identify a movie exists in IMDB but is missing in Douban, we will apply the model to the embedding feature from IMDB to infer its ranking score in Douban. Similarly, when building the {\our} model for IMDB, we will use the information in a reverse way: feature vector from Douban but ranking score from IMDB.

Formally, we can represent the ranking function involved in the {\our} model as
$${r}^{(1)}_i = \sigma(\mb{w}^\top \mb{y}^{(2)}_j + {b}),$$
where $\mb{w}$ and $b$ denote the weight and bias parameters of the ranking model. The loss introduced in the inference result ${r}^{(1)}_i$ and the ground-truth ranking score $\bar{r}^{(1)}_i$ can be represented as
$$\mathcal{L}_r(m^{(1)}_i) = ({r}^{(1)}_i - \bar{r}^{(1)}_i)^2.$$
Furthermore, given the Douban training movie set $\mathcal{M}^{(1)}$ the total loss introduced in building the model for Douban can be represented as 
\begin{align*}
\mathcal{L}_r(\mathcal{M}^{(1)}) &= \sum_{m^{(1)}_i \in \mathcal{M}^{(1)}} ({r}^{(1)}_i - \bar{r}^{(1)}_i)^2 = \left\| {\mb{r}}^{(1)} - \bar{\mb{r}}^{(1)} \right\|_2^2,
\end{align*}
where vectors ${\mb{r}}^{(1)}$ and $\bar{\mb{r}}^{(1)}$ contain the ranking scores of all movies in the training set. By minimizing the loss term, we can learn the parameters $\mb{w}$ and $b$ involved in the model. In a similar way, we can also train the missing movie ranking model for IMDB as well.


\section{Experiments}\label{sec:experiment}

\begin{figure}
\vspace{-25pt}
	\centering
	\includegraphics[width=0.4\textwidth]{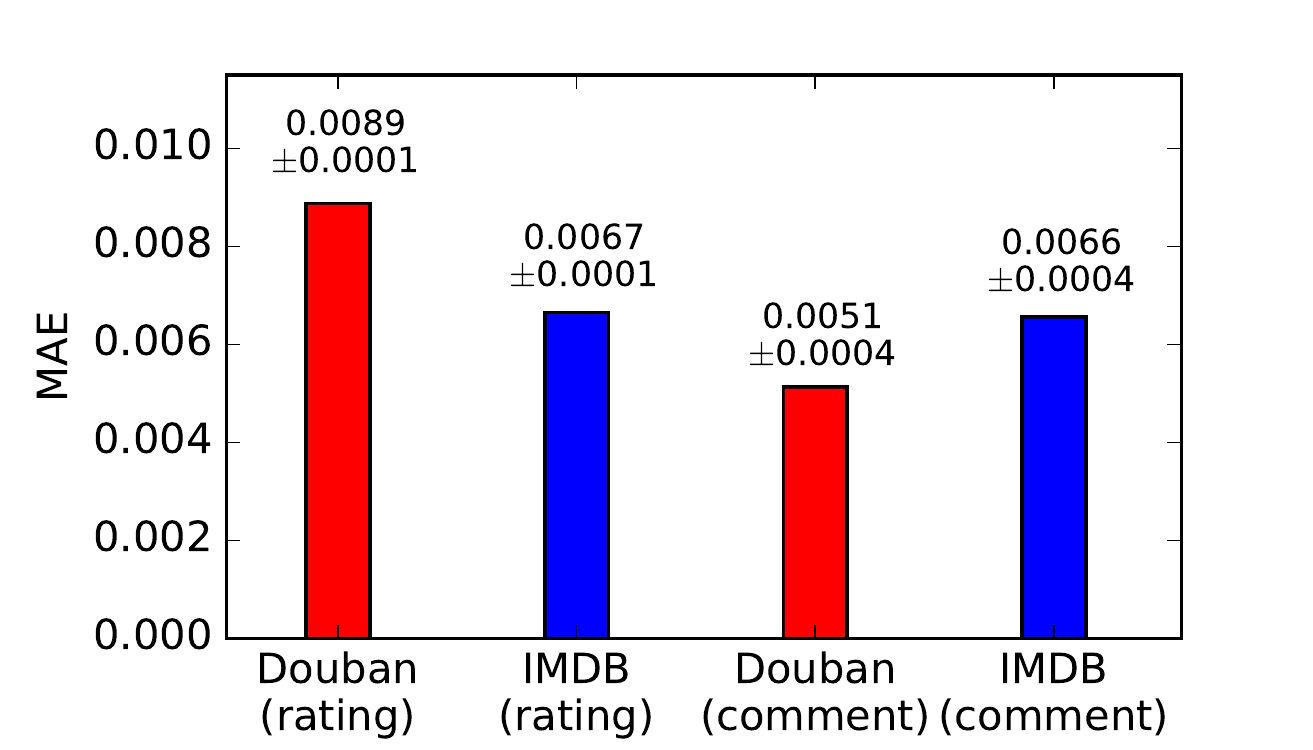}\vspace{-5pt}
	\caption{Ranking result of {\our} (rating: movie popularity rank, comment: movie quality rank).}
	\label{fig:mae}\vspace{-15pt}
\end{figure}

To test the effectiveness of the proposed model in identifying and ranking the missing movies, extensive experiments have been performed on the real-world aligned movie dataset introduced in Section~\ref{sec:general_analysis}. The experiments done on the Douban and IMDB datasets can be divided into two main parts: \textit{missing movie identification} and \textit{missing movie ranking}. Next, we will introduce the experimental settings and results of these two parts respectively.

\begin{table*}[t]
\vspace{-20pt}
\caption{Missing Movie Identification with various Missing Movie Ratios.}
\label{tab:missing_ratio_result}
\centering
{\tiny
\begin{tabular}{lrcccccccccc}
\toprule
\multicolumn{2}{l}{ }&\multicolumn{8}{c}{missing movie ratio $\delta$}\\
\cmidrule{3-12}
metric &method &0.1	& 0.2	& 0.3	& 0.4	&0.5	&0.6	&0.7	& 0.8	 &0.9	&1.0\\
\midrule
\multirow{5}{*}{\rotatebox{90}{Precision}}
&{\our} ($\eta$=1.0) &0.503$\pm$0.003 &0.504$\pm$0.003 &0.505$\pm$0.003 &0.507$\pm$0.004 &0.508$\pm$0.003 &0.509$\pm$0.003 &0.511$\pm$0.003 &0.512$\pm$0.004 &0.513$\pm$0.003 &0.513$\pm$0.003 \\
&{\our} ($\eta$=3.0) &0.535$\pm$0.042 &0.536$\pm$0.041 &0.537$\pm$0.041 &0.539$\pm$0.042 &0.54$\pm$0.041 &0.541$\pm$0.041 &0.543$\pm$0.041 &0.545$\pm$0.042 &0.545$\pm$0.041 &0.545$\pm$0.04 \\
\cmidrule{2-12}
&{\chinesematch} &\textbf{1.0}$\pm$\textbf{0.0} &\textbf{1.0}$\pm$\textbf{0.0} &\textbf{1.0}$\pm$\textbf{0.0} &\textbf{1.0}$\pm$\textbf{0.0} &\textbf{1.0}$\pm$\textbf{0.0} &\textbf{1.0}$\pm$\textbf{0.0} &\textbf{1.0}$\pm$\textbf{0.0} &\textbf{1.0}$\pm$\textbf{0.0} &\textbf{1.0}$\pm$\textbf{0.0} &\textbf{1.0}$\pm$\textbf{0.0} \\
&{\englishmatch} &\textbf{1.0}$\pm$\textbf{0.0} &\textbf{1.0}$\pm$\textbf{0.0} &\textbf{1.0}$\pm$\textbf{0.0} &\textbf{1.0}$\pm$\textbf{0.0} &\textbf{1.0}$\pm$\textbf{0.0} &\textbf{1.0}$\pm$\textbf{0.0} &\textbf{1.0}$\pm$\textbf{0.0} &\textbf{1.0}$\pm$\textbf{0.0} &\textbf{1.0}$\pm$\textbf{0.0} &\textbf{1.0}$\pm$\textbf{0.0} \\
\cmidrule{1-12}
\multirow{5}{*}{\rotatebox{90}{Recall}}
&{\our} ($\eta$=1.0) &\textbf{1.0}$\pm$\textbf{0.0} &\textbf{1.0}$\pm$\textbf{0.0} &\textbf{1.0}$\pm$\textbf{0.0} &\textbf{1.0}$\pm$\textbf{0.0} &\textbf{1.0}$\pm$\textbf{0.0} &\textbf{1.0}$\pm$\textbf{0.0} &\textbf{1.0}$\pm$\textbf{0.0} &\textbf{1.0}$\pm$\textbf{0.0} &\textbf{1.0}$\pm$\textbf{0.0} &\textbf{1.0}$\pm$\textbf{0.0} \\
&{\our} ($\eta$=3.0) &\textbf{0.99}$\pm$\textbf{0.02} &\textbf{0.99}$\pm$\textbf{0.02} &\textbf{0.99}$\pm$\textbf{0.02} &\textbf{0.99}$\pm$\textbf{0.02} &\textbf{0.99}$\pm$\textbf{0.02} &\textbf{0.99}$\pm$\textbf{0.02} &\textbf{0.99}$\pm$\textbf{0.02} &\textbf{0.99}$\pm$\textbf{0.02} &\textbf{0.99}$\pm$\textbf{0.02} &\textbf{0.99}$\pm$\textbf{0.02} \\
\cmidrule{2-12}
&{\chinesematch} &0.092$\pm$0.012 &0.092$\pm$0.012 &0.092$\pm$0.012 &0.092$\pm$0.012 &0.092$\pm$0.012 &0.092$\pm$0.012 &0.092$\pm$0.012 &0.092$\pm$0.012 &0.092$\pm$0.012 &0.092$\pm$0.012 \\
&{\englishmatch} &0.071$\pm$0.009 &0.071$\pm$0.009 &0.071$\pm$0.009 &0.071$\pm$0.009 &0.071$\pm$0.009 &0.071$\pm$0.009 &0.071$\pm$0.009 &0.071$\pm$0.009 &0.071$\pm$0.009 &0.071$\pm$0.009 \\
\cmidrule{1-12}
\multirow{5}{*}{\rotatebox{90}{F1}}
&{\our} ($\eta$=1.0) &\textbf{0.669}$\pm$\textbf{0.003} &\textbf{0.67}$\pm$\textbf{0.003} &\textbf{0.671}$\pm$\textbf{0.003} &\textbf{0.673}$\pm$\textbf{0.003} &\textbf{0.674}$\pm$\textbf{0.003} &\textbf{0.675}$\pm$\textbf{0.003} &\textbf{0.676}$\pm$\textbf{0.002} &\textbf{0.678}$\pm$\textbf{0.003} &\textbf{0.678}$\pm$\textbf{0.003} &\textbf{0.678}$\pm$\textbf{0.003} \\
&{\our} ($\eta$=3.0) &\textbf{0.693}$\pm$\textbf{0.03} &\textbf{0.694}$\pm$\textbf{0.029} &\textbf{0.695}$\pm$\textbf{0.029} &\textbf{0.696}$\pm$\textbf{0.029} &\textbf{0.698}$\pm$\textbf{0.028} &\textbf{0.698}$\pm$\textbf{0.028} &\textbf{0.7}$\pm$\textbf{0.029} &\textbf{0.702}$\pm$\textbf{0.029} &\textbf{0.702}$\pm$\textbf{0.028} &\textbf{0.702}$\pm$\textbf{0.028} \\
\cmidrule{2-12}
&{\chinesematch} &0.169$\pm$0.02 &0.169$\pm$0.02 &0.169$\pm$0.02 &0.169$\pm$0.02 &0.169$\pm$0.02 &0.169$\pm$0.02 &0.169$\pm$0.02 &0.169$\pm$0.02 &0.169$\pm$0.02 &0.169$\pm$0.02 \\
&{\englishmatch} &0.133$\pm$0.016 &0.133$\pm$0.016 &0.133$\pm$0.016 &0.133$\pm$0.016 &0.133$\pm$0.016 &0.133$\pm$0.016 &0.133$\pm$0.016 &0.133$\pm$0.016 &0.133$\pm$0.016 &0.133$\pm$0.016 \\
\cmidrule{1-12}
\multirow{5}{*}{\rotatebox{90}{Accuracy}}
&{\our} ($\eta$=1.0) &\textbf{0.501}$\pm$\textbf{0.001} &\textbf{0.502}$\pm$\textbf{0.001} &\textbf{0.504}$\pm$\textbf{0.001} &\textbf{0.505}$\pm$\textbf{0.002} &\textbf{0.507}$\pm$\textbf{0.002} &\textbf{0.507}$\pm$\textbf{0.002} &\textbf{0.509}$\pm$\textbf{0.002} &\textbf{0.511}$\pm$\textbf{0.002} &\textbf{0.511}$\pm$\textbf{0.002} &\textbf{0.512}$\pm$\textbf{0.003} \\
&{\our} ($\eta$=3.0) &\textbf{0.496}$\pm$\textbf{0.01} &\textbf{0.497}$\pm$\textbf{0.01} &\textbf{0.498}$\pm$\textbf{0.01} &\textbf{0.5}$\pm$\textbf{0.01} &\textbf{0.502}$\pm$\textbf{0.01} &\textbf{0.502}$\pm$\textbf{0.01} &\textbf{0.504}$\pm$\textbf{0.011} &\textbf{0.506}$\pm$\textbf{0.01} &\textbf{0.506}$\pm$\textbf{0.011} &\textbf{0.507}$\pm$\textbf{0.012} \\
\cmidrule{2-12}
&{\chinesematch} &0.046$\pm$0.006 &0.046$\pm$0.006 &0.046$\pm$0.006 &0.047$\pm$0.006 &0.047$\pm$0.006 &0.047$\pm$0.006 &0.047$\pm$0.006 &0.047$\pm$0.006 &0.047$\pm$0.006 &0.047$\pm$0.006 \\
&{\englishmatch} &0.036$\pm$0.005 &0.036$\pm$0.004 &0.036$\pm$0.005 &0.036$\pm$0.005 &0.036$\pm$0.005 &0.036$\pm$0.005 &0.036$\pm$0.005 &0.036$\pm$0.005 &0.036$\pm$0.005 &0.036$\pm$0.005 \\
\bottomrule

\end{tabular}\vspace{-14pt}
}
\end{table*}

\subsection{Experimental Setting: Missing Movie Identification}

The datasets used in the experiment include the Douban and IMDB OMKL datasets as introduced in Section~\ref{sec:general_analysis}. From the dataset, all the existing movie anchor links between Douban and IMDB are extracted and used as the positive set. A set of equal-sized non-existing movie anchor links are randomly sampled from the remaining Douban-IMDB movie pairs, which are used as the negative set. The positive and negative sets will be divided into two parts via the 10-fold cross validation, where 9 folds are used as the training set and 1 fold as the testing set.


Meanwhile, in the dataset, Douban and IMDB are almost fully aligned. To obtain a subset of the missing movies, a subset of the movies will be deleted from Douban (and IMDB). For instance, given three pairs of movies $\{(m^{(1)}_1, m^{(2)}_1), (m^{(1)}_2, m^{(2)}_2), (m^{(1)}_3, m^{(2)}_3)\}$ in Douban $G^{(1)}$ and IMDB $G^{(2)}$, by removing $m^{(1)}_2$ and $m^{(2)}_3$ from the pairs, the remaining movies $m^{(2)}_2$ and $m^{(1)}_3$ will be the missing movie for Douban and IMDB respectively. To control the number of missing movies, we use $\delta \in \{0.1, 0.2, \cdots, 1.0\}$ to represent the ratio of missing movies in the testing set. For instance, give the testing set $\mathcal{S}$, if $\delta = 0.9$, for all the positive movie anchor links in the testing set, we will preserve $10\%$ of them denoted as set $\mathcal{S}_1$ ($\mathcal{S}_1 \subset \mathcal{S}$, and $|\mathcal{S}_1| = 10\%|\mathcal{S}|$). Meanwhile, among the remaining $90\%$ of the positive anchor links (i.e., $\mathcal{S} \setminus \mathcal{S}_1$), we split them into two equal-sized disjoint subsets: $\mathcal{S}_2$ and $\mathcal{S}_3$ randomly. For all the positive anchor links in $\mathcal{S}_2$, we remove the Douban movies (and incident positive/negative links) from the dataset; and for all the positive anchor links in $\mathcal{S}_3$, we remove the IMDB movies (and incident positive/negative links) from the dataset. Therefore, for all the remaining movies involved in the set $\mathcal{S}_2$ and $\mathcal{S}_3$, they will be the missing movie for the other site.

Based on the training set, we aim at fitting the model introduced in the paper. For the movies in Douban and IMDB, a set of features are extracted, which include (1) movie profile information, like \textit{title}, \textit{length}, \textit{production year}, \textit{genre}, \textit{production country}, \textit{production language}, \textit{comment number}, \textit{average rating score}, \textit{storyline}; (2) movie cast information, including \textit{actor}, \textit{actress}, \textit{director}, and \textit{writer}; and (3) top 100 movie comments with the most likes. For the string information, like \textit{title}, \textit{storyline} and \textit{comments}, they are represented as the bag of words (in Chinese for Douban movies, and English for IMDB movies). The \textit{comment number}, \textit{average rating score}, \textit{length}, \textit{production year} of movies are represented as numerical features. As to the remaining information, they are represented in a similar way as the bag-of-words representation. For instance, we can represent all genres as the ``words'', if a movie belongs to certain genre, it will contain the ``genre word''. Similarly for \textit{country}, \textit{language}, \textit{actor}, \textit{actress}, \textit{director}, and \textit{writer}. These information in Douban and IMDB is in totally different representations. Meanwhile, the parameters involved in training the model are provided as follows: 3 hidden layers with neuron numbers $256$, $128$, $256$; epoch=50, batch size=512; weights $\alpha = 10.0$, $\beta = 0.01$ and $\gamma = 1000.0$.

For all the remaining links in the testing set, we can apply the learnt model to compute the embedded latent feature vector of the movies. For the Douban, IMDB movie pairs in the testing set, if their latent feature vectors are far away, they will be assigned with a negative label; otherwise, they will be assigned with a positive label. To measure the distance of the latent feature vector, we propose to apply the Euclidean distance. If the distance is smaller than threshold $\eta$, they will be labeled as the positive links; otherwise, they will be labeled as the negative links. The remaining nodes which are not connected by any anchor links will be identified as the missing movies.

\subsubsection{Comparison Methods}

Since the Douban and IMDB datasets are represented in different languages, few existing works have been done on the knowledge library mutual completion problem across multi-lingual sites. To show the advantages of the proposed model, we try to compare {\our} with several baseline methods based on title translation, where Google Translate\footnote{https://translate.google.com/} is used as the translation model.

\begin{itemize}

\item \textit{{\our}}: Model {\our} is the method proposed in this paper, which applies the deep aligned auto-encoder to obtain the latent feature vectors of the movies. For the movies which are relatively close with each other in the embedding space, they will be identified as the aligned movie pairs. 

\item \textit{Chinese Title Exact Matching ({\chinesematch})}: By translating the IMDB movie titles from English to Chinese, we can scan the movies in Douban to see whether there exist any matching movies sharing the same translated Chinese title. Exact Chinese word matching is applied here, and the matched movies will be paired together.

\item \textit{English Title Exact Matching ({\englishmatch})}: In a similar way, we can also translate the Douban movie title from Chinese to English to find the exact matching movies in IMDB. The method is called {\englishmatch} in the experiment.

\item \textit{Chinese Title Similarity ({\chinesescore})}: Exact matching of the translated movie titles occurs for a small number of movies only. However, in many other cases, the translated titles may share many common words, even though are not exactly matched. We propose to apply {\chinesescore} to calculate the similarity score of the movies by translating IMDB movie titles from English to Chinese, where the Chinese titles are represented as a set of words, and Jaccard's Coefficient is applied as the similarity measure here.

\item \textit{English Title Similarity ({\englishscore})}: We also apply the title similarity measures to the movies by translating the Douban movie titles from Chinese to English, and introduce the {\englishscore} baseline method.




\end{itemize}

\begin{figure*}[t]
\vspace{-20pt}
\centering
\subfigure[Precision]{ \label{fig:parameter_analysis_1}
    \begin{minipage}[l]{0.45\columnwidth}
      \centering
      \includegraphics[width=1.0\textwidth]{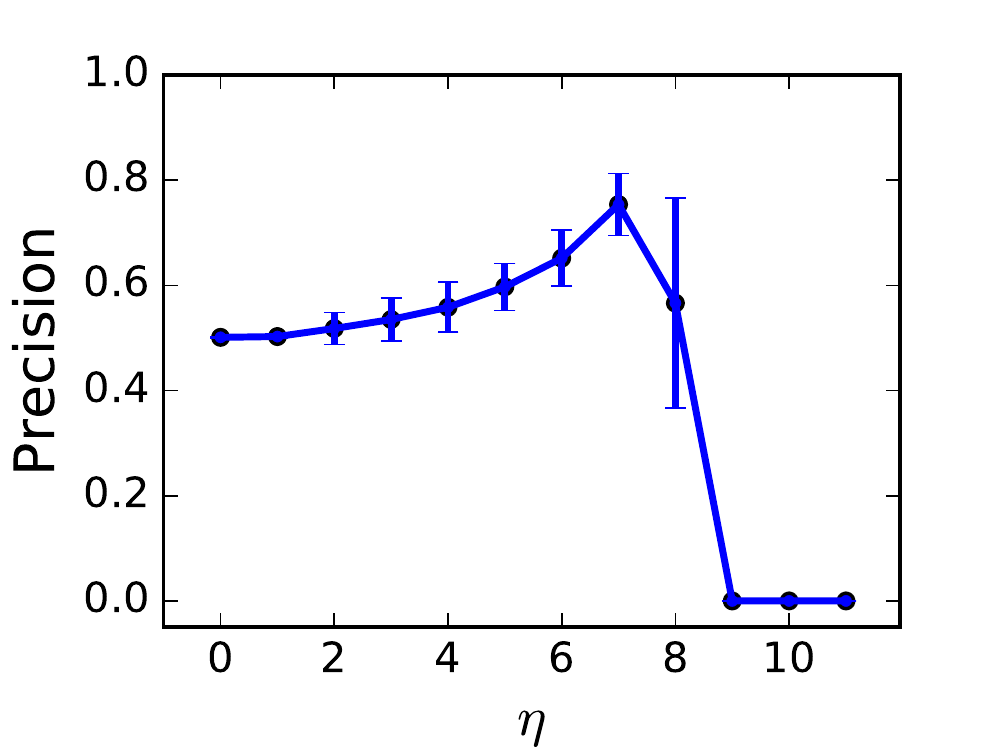}
    \end{minipage}
}
\subfigure[Recall]{ \label{fig:parameter_analysis_2}
    \begin{minipage}[l]{0.45\columnwidth}
      \centering
      \includegraphics[width=1.0\textwidth]{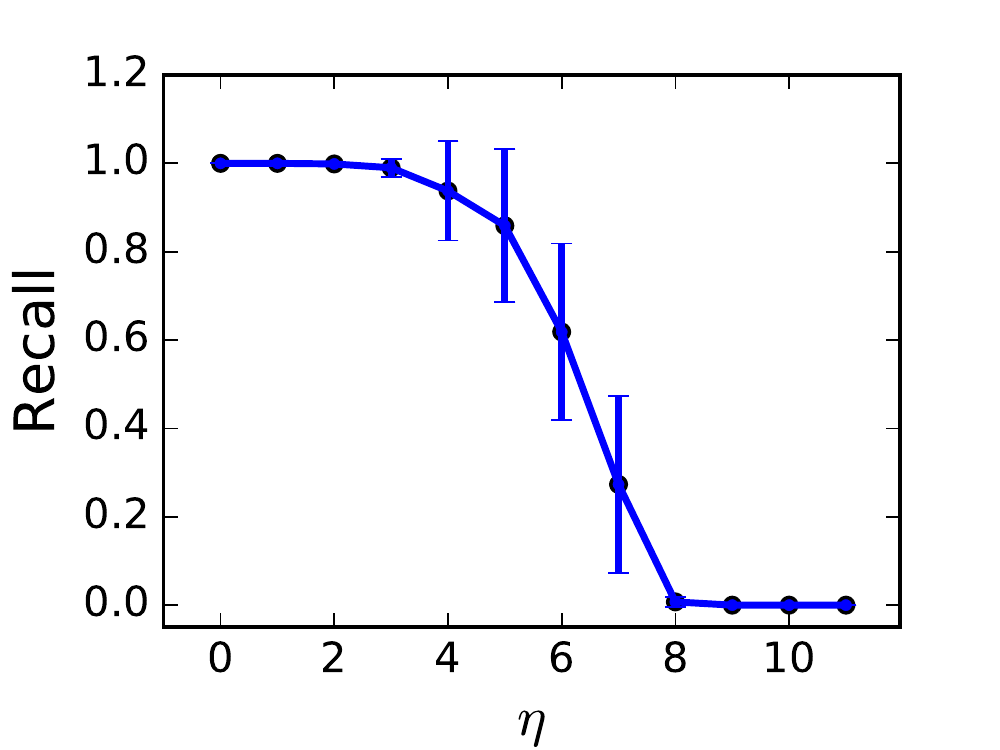}
    \end{minipage}
}
\subfigure[F1]{\label{fig:parameter_analysis_3}
    \begin{minipage}[l]{0.45\columnwidth}
      \centering
      \includegraphics[width=1.0\textwidth]{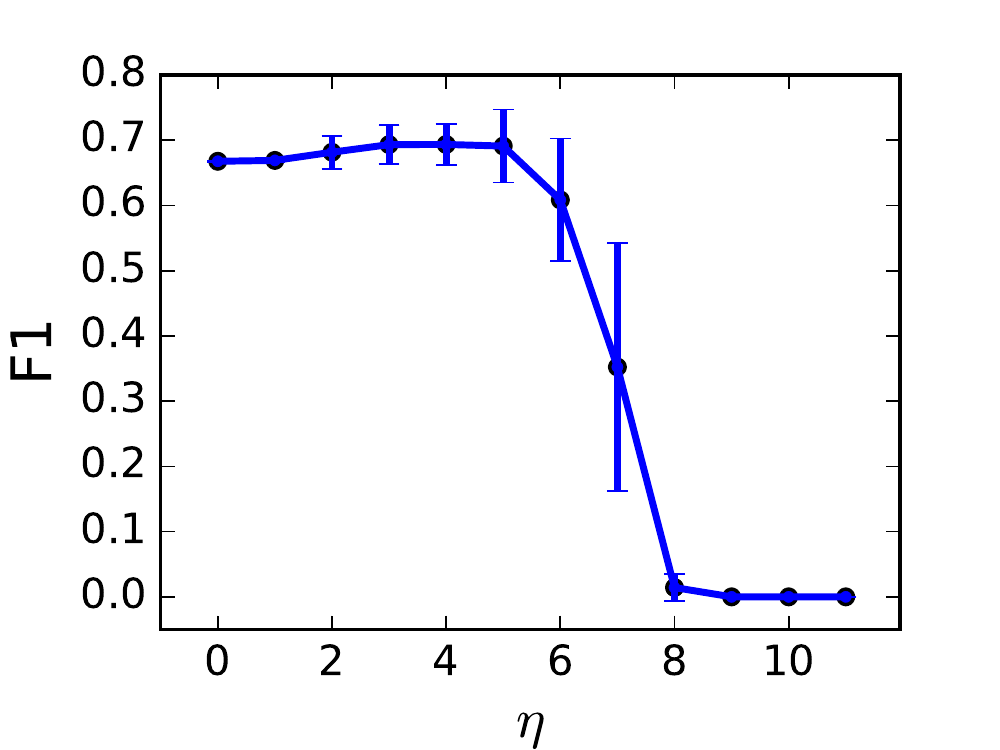}
    \end{minipage}
}
\subfigure[Accuracy]{ \label{ffig:parameter_analysis_4}
    \begin{minipage}[l]{0.45\columnwidth}
      \centering
      \includegraphics[width=1.0\textwidth]{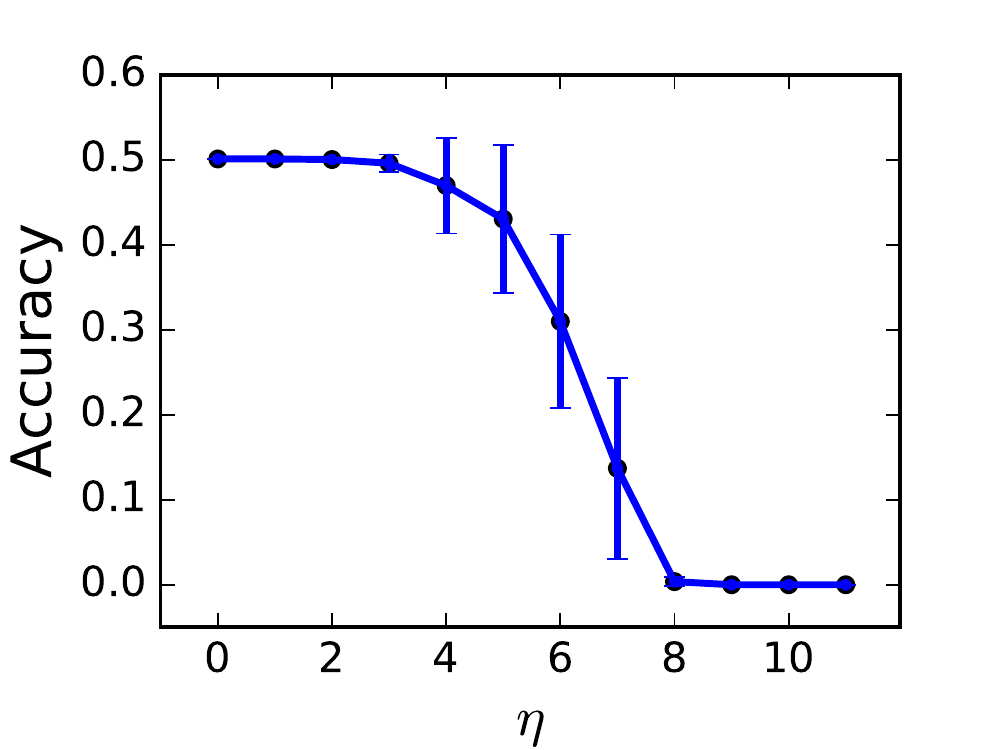}
    \end{minipage}
}\vspace{-5pt}
\caption{Parameter Analysis of $\eta$.}\label{fig:parameter_analysis}\vspace{-15pt}
\end{figure*}

\subsubsection{Evaluation Metrics}

To evaluate the performance of {\our}, {\chinesescore} and {\englishscore}, we will apply AUC, Precision@K (at top K) as the evaluation metrics. With the distance threshold $\eta$, we can also determine the labels of the movie pairs for method {\our}. Furthermore, we also compare the performance of {\our}, {\chinesematch} and {\englishmatch} by applying Accuracy, Precision, Recall and F1 as the evaluation metrics.

\subsection{Experimental Result: Missing Movie Identification}

The experimental results for the missing movie identification is available in Figure~\ref{fig:anchor_link} and Table~\ref{tab:missing_ratio_result}, where the missing ratio $\delta$ changes with value in $\{0.1, 0.2, \cdots, 1.0\}$.

According to the results in Figure~\ref{fig:anchor_link} , framework {\our} proposed in this paper can outperform the translation based baseline methods consistently with different missing movie ratio $\delta$s in inferring the movie anchor links. As $\delta$ changes, the performance of {\our} doesn't change much, which denotes that {\our} is not sensitive to parameter $\delta$, and is applicable to the OMKLs with various percentage of the missing movies. For instance, when $\delta = 0.1$, the AUC achieved by {\our} is $0.881$, which $13.5\%$ larger than the AUC achieved by {\chinesescore}, and over $30\%$ larger than the AUC obtained by {\englishscore}. The advantage of {\our} compared with {\chinesescore} and {\englishscore} will be more significant evaluated the Prec@200 metric. The Prec@200 obtained by {\our} is almost the double of those obtained by the other two comparison methods. 

With threshold $\eta$, {\our} will assign the movie pairs whose distance is smaller than $\eta$ with the $+1$ label. In Table~\ref{tab:missing_ratio_result}, we compare {\our}  ($\eta = 1.0$ and $\eta = 3.0$) with the exact title matching based methods {\chinesematch} and {\englishmatch}. With exact matching, the movie anchor links identified by {\chinesematch} and {\englishmatch} are all correct. Therefore, {\chinesematch} and {\englishmatch} can achieve $1.0$ Precision score for different missing movie ratio $\delta$. However, exact title matching may miss many movie anchor links as well, and the Recall scores obtained by these two methods are extremely low. Meanwhile, {\our} can correctly predict all the movie anchor links, which can achieve $1.0$ Recall, as well as very good Precision score at the same time. By considering Precision and Recall simultaneously, {\our} can outperform {\chinesematch} and {\englishmatch} with very big advantages evaluated by F1. For instance, the F1 score obtained by {\our} is almost the triple of the score obtained by {\chinesematch} and {\englishmatch}. Similar results can be observed for the Accuracy metric.

As we can see, with different threshold $\eta$, {\our} will have different performance according to the results in Table~\ref{tab:missing_ratio_result}. To study the effects of $\eta$ on {\our}, we also carry out a sensitivity analysis of $\eta$ and the results are provided in Figure~\ref{fig:parameter_analysis}, where $\eta$ changes with values in $\{0.1, 1.0, 2.0, \cdots, 11.0\}$. According to the plot, as $\eta$ increases, the performance of {\our} will slightly improve first and then drop quickly. With $\eta = 0.1$, {\our} can already obtained very food performance, which shows that {\our} can effectively project the aligned movie pairs to close regions with a distance less than $0.1$. Meanwhile, the performance of {\our} degrades until $\eta$ increases to $4$ for Recall, F1 and Accuracy (or $7$ for Precision). It shows that {\our} can also project the non-anchor movie pairs to different regions, whose distance (i.e., the $L_2$ norm of their embedding vector difference) will be generally greater than $4.0$.

\subsection{Experimental Setting: Missing Movie Ranking}

The experimental setting for the movie ranking step is very similar to the previous missing movie identification setting. Based on the training and testing set obtained via cross validation in the previous step, we can obtain the latent feature vectors of all the movies connected by the positive movie anchor links respectively. Based on the latent feature vector of the Douban movies involved in the positive training instances and the ranking scores of their corresponding IMDB movies (i.e., the movies they connected with via the anchor links), we can build a ranking model as introduced in the paper. For the identified missing movies of IMDB, we will apply the model to the latent feature vectors of the Douban movies, to infer their potential scores to be obtained in the IMDB site. For these missing movies of IMDB, we can obtain their ground-truth ranking scores in IMDB, which will be used for evaluation purposes. Similarly, we will also build the ranking models for the Douban site. Depending on the specific ranking criteria being used, the ranking model will project the movies to different types of ranking scores, i.e., the \textit{movie quality rank} and \textit{movie popularity rank} respectively.

The evaluation metrics used for the missing movie ranking experiments include MAE (mean square error).

\subsection{Experimental Result: Missing Movie Ranking}

The missing movie ranking are based on the identified missing movies from the previous step, whose results are provided in Figure~\ref{fig:mae}. In Figure~\ref{fig:mae}, we show the ranking results for missing movies in Douban and IMDB based on the \textit{movie quality rank} and \textit{movie popularity rank} respectively, whose performance is evaluated by the MAE metric. According to the result, {\our} can learn the ranking scores of the missing movies very well, whose errors are much lower than $0.01$ for all these movies. For instance, the MAE obtained by {\our} for inferring the \textit{movie popularity rank} in Douban is $0.0051$, which is a very good small inference error.

\section{Related Works}
\label{sec:related_works}

Until recently, very few works have been done on cross-language content creation. To the best of our knowledge, the most related work to this paper is \cite{WWZL16}, which proposes to grow the Wikipedia website by adding new articles based on the Wikipedia websites in other countries. Wulczyn et al. \cite{WWZL16} carries out a systematic analysis of the existing Wikipedia datasets, and propose to identify the missing articles from the perspective of the language-independent concepts. Some other works have been done on the entity resolution problem among different sites, like the Wikipedia articles and the patents \cite{YSTML15}. Yang et al. \cite{YSTML15} study the problem from the language topic level instead, by extracting the topics of Wikipedia and patent articles, a matching task is performed to learn the mapping relationships of articles from different sources. Hale et al. \cite{H13} propose to analyze one month of edits to Wikipedia and examine the role of users editing multiple language editions (i.e., multilingual users). According to the discovery in \cite{H13}, multilingual users are much more active than their single-edition (monolingual) counterparts and are found in all language editions, but smaller-sized editions with fewer users have a higher percentage of multilingual users than larger-sized editions. Adar et al. \cite{ASW09} introduce an automated system for aligning Wikipedia infoboxes, creating new infoboxes as necessary, filling in missing information, and detecting discrepancies between parallel pages.

From the language processing study perspective, some works have been done to group the entities about the same entities across languages. Green et al. \cite{GAG12} propose to cluster the text mentions across different languages, which may refer to the same concept. They test their model based on the Arabic-English corpus and yields a good performance. Lawrie et al. \cite{LMMO14} propose to study the problem of entity linking to associate references to some entity that are found in unstructured natural language content to an authoritative inventory of known entities. Mayfield et al. \cite{MLMO11} introduce an efficient way to create a test collection for evaluating the accuracy of cross-language entity linking. They \cite{MLMO11} apply the technique to produce the first publicly available multilingual cross-language entity linking collection, which includes approximately 55,000 queries, comprising between 875 and 4,329 queries for each of twenty-one non-English languages.

Other closely related problems to the task studied in this paper include network alignment and network matching. Network alignment problem has also been studied by many works, which was mostly applied to address many applications in bioinformatics, e.g., protein-protein interaction (PPI) network alignment \cite{Klau09, SXB09, SXB07, KMMHP10, K09, LLBSB09}. Most network alignment approaches focus on finding approximate isomorphism between two graphs under unsupervised settings, e.g., \cite{SXB07, KMMHP10, K09}. Because the intractability of the problem, existing methods usually rely on practical heuristics to solve the alignment problem \cite{K09, LLBSB09}. Different from these works, Bayati et al. \cite{BGGSW09} proposed to use belief propagation to solve sparse network alignment problems and Todor proposes a probabilistic biological network alignment algorithm in \cite{TDK13}. Biological networks can be of very large sizes and many scalable alignment methods have been proposed in \cite{SP11, KBS08, SY13, AE13}.

Meanwhile, in recent years, some works have been done on aligning social networks \cite{KZY13, KTL13, LSY14}. To prune the predicted anchor links in network alignment problems, stable matching is first applied by the PI in \cite{KZY13}, while assuming if a person has multiple accounts in a social network, these accounts can be identified and consolidated as a single account through a preprocessing step \cite{LWC12}. College admission problem \cite{R85} and stable marriage problem \cite{GI89} have been studied for many years and lots of works have been done in the last century. In recent years, some new papers have come out in these areas. Sotomayor et al. \cite{S08} propose to analyze the stability of the equilibrium outcomes in the admission games induced by stable matching rules. Meanwhile, in college admission problems, only the small core in nash equilibrium plays an important role, which was analyzed in \cite{M98}. Flor\'{e}en et al. \cite{FKPS10} propose to study the almost stable matching by truncating the Gale-Shapley algorithm. In this proposal, we will address the \textit{partial network alignment} problem by extending traditional stable matching to a generalized case to identify the non-anchor users and prune the redundant anchor links connected to them.

\section{Conclusion}\label{sec:conclusion}

In this paper, we have studied the {\problem} to identify and rank the missing movies for the completion of \textit{multiple isomeric online knowledge libraries}. Several new concepts have been clearly defined in the paper, which include \textit{online movie knowledge library} (OMKL), \textit{multiple aligned isomeric OMKLs}, and the \textit{movie anchor links}. A thorough analysis of two aligned \textit{aligned isomeric OMKLs} (i.e., Douban and IMDB) has been carried out in the paper, which provides useful insights for building the framework {\our} to address the {\problem} problem. Extensive experiments done on Douban and IMDB real-world datasets have demonstrated the effectiveness of {\our} in identifying and ranking the missing movies.
\section{Acknowledgement}\label{sec:ack}

This work is supported in part by NSF through grants IIS-1526499, IIS-1763325, and CNS-1626432, and NSFC 61672313. This work is also partially supported by NSF through grant IIS-1763365 and by FSU through the startup package and FYAP award.
\balance
\bibliographystyle{plain}
\bibliography{reference}

\end{document}